%% file: aanda.tex
\documentclass[longauth]{aa} % for the long lists of affiliations 
\usepackage{graphicx}
\usepackage{textcomp}
\usepackage{txfonts}
\usepackage[dvipsnames]{xcolor}

\AddToHook{begindocument/before}{\RequirePackage[colorlinks=true,
linkcolor=blue, citecolor=blue, filecolor=blue, urlcolor=blue,
unicode=true]{hyperref}}
\usepackage{soul}
\usepackage{lipsum}
\usepackage{amsmath}

\def\ts     {\thinspace}

\def\kms  {\ifmmode{{\rm \ts km\ts s}^{-1}}\else{\ts km\ts s$^{-1}$\ts}\fi}
\def\msol {\ifmmode{{\rm M}_{\odot}}\else{M$_{\odot}$\ts}\fi}
\def\lsun {\ifmmode{{\rm L}_{\odot}}\else{L$_{\odot}$\ts}\fi}
\def\cii  {\ifmmode{{\rm [C}{\rm \scriptstyle II}]}\else{[C\ts {\scriptsize II}]\ts}\fi}
\def\ciii  {\ifmmode{{\rm [C}{\rm \scriptstyle III}]}\else{[C\ts {\scriptsize III}]\ts}\fi}
\def\ci   {\ifmmode{{\rm C}{\rm \scriptstyle I}}\else{C\ts {\scriptsize I}\ts}\fi}
\def\m    {\ifmmode{\mu {\rm m}}\else{$\mu$m}\fi}
\def\hi   {\ifmmode{{\rm H}{\rm \scriptstyle I}}\else{H\ts {\scriptsize I}\ts}\fi}
\def\hii  {\ifmmode{{\rm H}{\rm \scriptstyle II}}\else{H\ts {\scriptsize II}\ts}\fi}
\def\nii  {\ifmmode{{\rm [N}{\rm \scriptstyle II}]}\else{[N\ts {\scriptsize II}]\ts}\fi}
\def\oiii {\ifmmode{{\rm [O}{\rm \scriptstyle III}]}\else{[O\ts {\scriptsize III}]\ts}\fi}
\def\hh   {\ifmmode{{\rm H}_2}\else{H$_2$\ts}\fi}
\def\nhh  {\ifmmode{N({\rm H}_2)}\else{$N$(H$_2$)\ts}\fi}
\def\microns {\ifmmode{\mu{\rm m}}\else{$\mu$m\ts}\fi}

\definecolor{lightblue}{HTML}{d7e5ff}

\def\source{\ifmmode{{\rm CRISTAL-05}}\else{CRISTAL-05\ts}\fi}
\begin{document}

%\AddToHook{begindocument/before}

%\begin{document}
%begindocument/before

   \title{The ALMA-CRISTAL survey: Extended [CII]  emission in an interacting galaxy system at z $\sim$ 5.5}

   \subtitle{} %CRISTAL survey: merger as the cause of \cii halos. CRISTAL survey: 

    \input{authors}

  \abstract
   {The ALMA \cii Resolved Ism in STar-forming gALaxies (CRISTAL) survey is a Cycle 8 ALMA Large Programme that studies the cold gas component of high-redshift galaxies. Its sub-arcsecond resolution observations are key to disentangling physical mechanisms that shape galaxies during cosmic dawn.
   In this paper, we explore the morphology and kinematics of the cold gas, star-forming, and stellar components in the star-forming main-sequence galaxy CRISTAL-05/HZ3, at z = 5.54. Our analysis includes 0.3\arcsec $\,$ spatial resolution ($\sim$ 2 kpc) ALMA observations of the \cii line. 
   While CRISTAL-05 was previously classified as a single source, our observations reveal that the system is a close interacting pair surrounded by an extended component of carbon-enriched gas. This is imprinted in the disturbed elongated \cii morphology and the separation of the two components in the position-velocity diagram ($\sim 100$ \kms). The central region is composed of two components, named C05-NW and C05-SE, with the former being the dominant one. A significant fraction of the \cii arises beyond the close pair up to 10 kpc, while the regions forming new massive stars and the stellar component seem compact (r$_{\cii} \sim 4 \times$ r$_{UV}$), as traced by rest-frame UV and optical imaging obtained with the Hubble Space Telescope and the James Webb Space Telescope. Our kinematic model, constructed using the DYSMALpy software, yields a minor contribution of dark matter of C05-NW within a radius of $\sim$ 2 $\times$ R$\rm_{eff}$. Finally, we explore the resolved \cii/FIR ratios as a proxy for shock-heating produced by this merger. We argue that the extended \cii emission is mainly caused by the merger of the galaxies, which could not be discerned with lower-resolution observations. Our work emphasizes the need for high-resolution observations to fully characterize the dynamic stages of infant galaxies and the physical mechanisms that drive the metal enrichment of the circumgalactic medium.}

    %{}

    %{}


   \titlerunning{\cii extended emission in an interacting galaxy system at z $\sim$ 5.5}
   \maketitle
    
%
%-------------------------------------------------------------------

% \input{pre-introduction-note}

\section{Introduction}

Galaxy growth is governed by an interplay of several physical processes that transform the galactic internal properties across cosmic time \citep{2015somerville,2017tumlinson, 2017naab}. The current galaxy evolution paradigm suggests that gas accretion from the intergalactic medium and intense merger activity provide cold gas to fuel star formation in a galaxy, while active galactic nuclei (AGN) and stellar feedback can deplete and expel the internal gas outwards the interstellar medium (ISM), eventually halting star-formation \citep{2005berry, 1991barnes,1996sanders,2012lambas, 2014almeida, 2016hayward, 2017harrison}. We are in the first steps of detecting and characterizing the full baryonic cycle at the early stages of galaxy formation. 

Since these intricate physical mechanisms shape the spatial distribution of the baryonic matter, the morphology becomes the primary indicator of the evolutionary history of these systems. For example, when dynamical events occur, such as mergers and outflows, they can get disrupted and distorted \citep{2015somerville, 2019nelson, 2019pillepich}. Furthermore, the diversity of shapes of low-redshift galaxies correlates with several physical properties, such as the stellar mass, star formation histories, and environment \citep{2004baldry, 2006delucia,2010poggianti, 2013tojeiro,2021villanueva}. For instance, elliptical and lenticular galaxies reside preferentially in high-density environments, populating especially the cores of clusters and groups, while spiral galaxies inhabit field regions \citep{1980dressler, 2005postman}. However, high-redshift galaxies (i.e., distant) are fainter, have a smaller angular size, and suffer from cosmological dimming which makes detecting subtle features challenging, and thus the physical mechanisms that had shaped them. 
    
Most spatially-resolved studies of high-redshift galaxies rely on rest-UV imaging from the Hubble Space Telescope (HST), as it was the only instrument that provides sufficiently high spatial resolution and depth to reveal the inner structures of small galaxies at these wavelengths. Such observations possibly suggest that galaxies at $z > 4$ are clumpy and irregular, indicating high merger activity in the first billion years of the Universe \citep{2009conselice,2019duncan,2022shibuya}. However, the HST imaging may not provide a full answer because the rest-UV emission is strongly affected by dust obscuration and biased towards the most intense ongoing star-forming sites. Recent rest-frame near-infrared James Webb Space Telescope (JWST) imaging has revealed that many of these clumps are just the compact star-forming regions of disk galaxies, which surprisingly became a common population ($\sim$ 50\%) at z = 3-6 \citep{2022ferreira,2023guo,2023kartaltepe,2023huertas-company}.

%% Minor edits by MA
A complete picture of the evolutionary path of galaxies inevitably requires information on the cold gas supply necessary to form stars. The start of operation of the Atacama Large Millimeter/submillimeter Array
(ALMA) in the last decade was the pivotal point to a new era in assessing the gas at high-redshift in sub-arcsec scales. The cornerstone of these groundbreaking studies has been the \cii emission line, arising from the fine-structure transition $^{2}{\rm P}_{3/2}$ → $^{2}{\rm P}_{1/2}$ of the singly ionized carbon atom (hereafter just \cii, centered at 158 $\mu$m). This is one of the strongest far-infrared (FIR) lines, reaching up to 1\% of the total bolometric luminosity \citep{2010stacey,2013diazsantos,2015cormier,2018herreracamus, 2019rybak,2020schaerer}. Due to its low ionization potential of 11.2 eV, it is a dominant gas coolant of the neutral ISM coming from photo-dissociation regions (PDR), cold molecular, ionized, and atomic gas phases  \citep{2003wolfire, 2015vallini, 2019clark}.

%% by MA
A remarkable outcome of recent studies on the spatial distribution of the gas- and star-forming components of galaxies at $z=4-6$ is the existence of extended \cii line emission reaching out to a radius of $\sim$10 kpc, or about 5$\times$ beyond the extent of the star-forming areas traced by rest-UV and dust continua \citep{2019fujimoto}. These so-called ``\cii halos'' were first identified by \cite{2019fujimoto} through the radial profiles of stacked \cii emission of 18 normal (L$_{UV} >$ 0.6 L$^{*}$) star-forming galaxies at $z>4$, selected as single systems with no obvious AGN signatures. Several studies have found similar \cii structures in individual star-forming galaxies at $z>4$ \citep{2020fujimoto, 2021herrera-camus, 2022akins, 2023lambert} and massive quasar host galaxies at $z>6$ \citep{2012maiolino,2015cicone,2022meyer}. \cii halos appear to be ubiquitous in the most massive galaxies yet less relevant in lower mass systems (M$_{*} \lesssim 10^{10} \rm \, M_{\odot}$) \citep{2020fujimoto, 2023posses, 2023pizzati}.

The presence of emitting carbon-rich gas beyond galactic sizes raises the question of how enriched gas was transported to larger distances whilst remaining in a singly ionized state. Mock observations based on state-of-the-art cosmological simulations fail to reproduce the \cii surface brightness of observed galaxies \citep{2019fujimoto,2019pallotini,2020arata}. Several physical mechanisms for the origin of these `halos' have been proposed in the literature,  including the presence of unresolved satellite galaxies around the central source, extended photodissociation or \hii regions, cold streams, and outflows \citep{2019fujimoto, 2024dicesare}. Several studies have detected broad components in the \cii spectrum of stacked and individual galaxies, suggesting that the \cii halos are the result of past/ongoing episodes of outflowing gas driven by powerful stellar feedback \citep{2018gallerani,2020ginolfia, 2021herrera-camus}. This is supported by recent semi-analytical models of catastrophically cooling outflows ($T \approx 10^{2-4}$ K), which can reproduce the observed \cii surface brightness profiles \citep{2020pizzati,2023pizzati}.

An additional scenario was proposed by \cite{2020ginolfib}, who identified a luminous envelope of \cii emission surrounding a dense merging system at z $\sim$ 4.5. This phenomenon leads to tidal stripping, where the gas is removed from the ISM and enriches the circumgalactic medium (CGM) with chemically enhanced material. Similarly, \citet{2023lambert} reported a \cii extended emission in a typical galaxy at z = 5.25, showing indications of a late-stage merger rather than outflow signatures. Most studies rely on observations with a resolution of $\sim1\arcsec$ ($\sim$5 kpc), which can resolve the current 10-kpc halos in less than two resolution units. The understanding of the dynamic state of these systems that can support or rule out the last scenario is also constrained by the resolution and depth of the observation since they impact the determination of the nature of single/multiple sources \citep{2022rizzo, 2020forster}. 

In this context, the ALMA large program CRISTAL (\cii Resolved Ism in STar-forming gALaxies) survey provides the resolution and depth needed to constrain the critical conditions to solve these open questions (2021.1.00280.L; Herrera-Camus et al. in prep.). CRISTAL aims to dissect the \cii and dust continuum in 25 main-sequence star-forming galaxies at z = 4 - 6 with available ALMA lower-resolution observations and UV continuum detections. The CRISTAL observations are more sensitive and have higher spatial resolution than previous observations, resolving these galaxies at 1-2 kpc scales, which is crucial to confirm the detection and identify the physical mechanisms powering the emission at such distances. This paper, as part of the following series of CRISTAL results \citep{2023Mitsuhashi, 2024solimano}, reports our analysis of the compelling case of the main-sequence star-forming galaxy HZ3, identified in the survey as CRISTAL-05, at z = 5.54 \citep{2015capak}. This was one of the galaxies reported by \citet{2020fujimoto} to host a \cii halo, but yet with an uncertain description of the spatial and dynamical internal structure. Here, we take advantage of the deepest ALMA integration in the CRISTAL sample (8.8 $\mu$Jy/beam), owing to previous deep pilot observations, to study the nature of the system and the origin of the puzzling \cii halo. Throughout this article, we m refer to a \cii halo as extended \cii emission. As discussed in \citet{2023lambert}, this nomenclature has become ubiquitous throughout the literature. But at the same time, this is an unusual use of the term halo since the well-establish halo component (dark matter haloes, stellar haloes, and galactic haloes) refers to a spheroidal shape that we can not conclude with the resolution in hand.

This paper is laid out as follows. Firstly, we describe the general properties of the galaxy, as well as the ALMA observations and ancillary data in Section 2. In Section 3, we detail the main results: we characterize the morphology of the \cii emission and compare it to the rest-frame UV and dust continuum. We also quantify the multiwavelength morphology and sizes. The gas motions are then determined by the moment-1 map, moment-2  map, and position-velocity diagram. After that, we dissect the substructure of the system in different components with a clump-finding algorithm. We finally perform a kinematical disk modeling in one of the components of CRISTAL-05. Section 4 presents a discussion of the results of the previous section. Lastly, the paper ends with Section 5, a summary and main conclusions of the work. We assume a standard, flat $\Lambda$CDM model with: $\Omega_{\Lambda}=0.7$, $\Omega_{\rm M}=0.3$ and ${\rm H_{0}}=70\,{\rm km}\,{\rm s}^{-1}\, {\rm Mpc}^{-1}$, which implies a physical factor of 5.961 kpc/\arcsec at the redshift of the source.

%--------------------------------------------------------------------

\section{Observations and Data Reduction}

A brief description of the target and all datasets analyzed in this work is presented below.

\begin{table}
\centering       
\begin{tabular}{l c c c}     % 7 columns 
\hline\hline       
                      % To combine 4 columns into a single one 
Product                & robust   &  rms                  & Beam   \\ 
                       &          &  $\mu$Jy beam$^{-1}$  &             \\
\hline            
   \\
    Continuum          & 2        &  8.8                  &   0.33\arcsec $\times$ 0.27\arcsec     \\ 
    \\ 
    Cube 10 \kms              & 2        &  150                   &  0.36\arcsec $\times$ 0.29\arcsec     \\ 
    Cube 20 \kms               & 2        &  123                   &  0.31\arcsec    $\times$ 0.26\arcsec \\ 

    \\ 

    \cii map           & 0.5     &  45                    &  0.21\arcsec $\times$ 0.18\arcsec    \\ 
    \cii map           & 2       &  42                    &  0.33\arcsec    $\times$ 0.27\arcsec    \\ 

   \\ 

\hline  
\hline

\end{tabular}
\caption{Properties of the main ALMA products. The velocity associated with each cube refers to the velocity resolution of the channels.}
\label{tab:products}  
\end{table}

%\begin{table}
%\caption{Observational and physical properties of \target}
%\centering          
%\begin{tabular}{l c }     % 7 columns 
%\hline\hline       
%                      % To combine 4 columns into a single one 
%Property & Value \\ 
%\hline            
%   \\
%   Center                                         & +10:00:29.86, +02:13:02.19    \\  
%   $z_{\mathrm{[CII]}}$$^{a}$                     &  5.5420 $\pm$ 0.0005      \\ 
%   SFR$_{\rm UV}$ (M$_{\odot}$ yr$^{-1}$)         &   22 $\pm$ 1                    \\ 
% 
%   SFR$_{\rm IR}$$^{b}$ (M$_{\odot}$ yr$^{-1}$)   &  1.38 $\pm$ 0.09   \\ 
%   Stellar mass$^{b}$ ($10^{10}$ M$_{\odot}$)     &   1.4 $\pm$ 0.6               \\ 
%    
%   \\
%   S/N$^{c}$ (\cii)                                     & 14                   \\ 
%   
%   \cii line flux (Jy km s$^{-1}$)                & 1.00 $\pm$ 0.07                \\  
%   FWHM$_{\rm [CII]}$ (km s$^{-1}$)               & 238  $\pm$ 25                   \\ 
%   S$_{\rm 158\mu m}$$^{b}$        (mJy)                &  0.14 $\pm$ 0.04          \\ 
%   L$_{\rm [CII]}$ ($10^{8}$ L$_{\odot}$)         & 9.3  $\pm$ 0.7                  \\ 
%   L$_{\rm IR}$$^{b}$  ($10^{11}$ L$_{\odot}$)           &  1.60  $\pm$ 0.08                \\ 
%   \\ %3.6 $\pm$ 0.5
%
%
%  % &  &  \\ 
%   %&  &  \\ 
%   %&  &  \\ 
%
%\hline  
%\hline
%
%\end{tabular}
%\flushleft{{\bf Notes:} Values derived in this work were obtained in a 1.5\arcsec-radius aperture. $^{a}$ \citet{2020bethermin}, $^{b}$Measured by \citet{2023Mitsuhashi}, $^{c}$Integrated S/N. } 
%\label{tab:properties-central}  
%\end{table}

\subsection{Target: CRISTAL-05}

%, and a non-detection of dust continuum emission, indicating that the galaxy has a low dust content \citep{2015capak,2017barisic, 2017faisst}. 
CRISTAL-05 \citep[also known as HZ3 and COSMOS2015 683613,][]{2015capak,2016laigle} was identified as a  Lyman break galaxy candidate at z $\sim$ 5 in the Cosmic Evolution Survey \citep[COSMOS,][]{2016laigle}, and spectroscopically confirmed based on UV absorption lines \citep[][]{2013ilbert}. ALMA follow-up revealed bright \cii emission at ${\rm z}_{\cii} = 5.546 \pm 0.004$ \citep{2015capak}. Its UV-optical derived stellar mass and star formation rate (SFR) place CRISTAL-05 on the "main sequence" relationship at z $\sim$ 5.5 \citep[e.g.,][]{2014speagle}, and with L$_{UV}$ $\sim$ L$^{*}$. As a typical star-forming galaxy at its redshift, it was thus included as part of the  ALMA Large Program to Investigate C+ at Early Times survey \citep[ALPINE, ][]{2020lefevre,2020bethermin,2020faisst}. \citet{2020lefevre} classified it as an extended dispersion-dominated system, based on a morpho-kinematic analysis, but a more detailed analysis by \citet{2021jones} updated it to an uncertain classification as it revealed a potential blue,
compact source in the south portion of the galaxy. \citet{2018carniani} found that while the galaxy deviates positively from the local L$_{\cii}$-SFR relation, it presents a \cii deficiency in the $\Sigma_{\cii}$ - $\Sigma_{\rm SFR}$ parameter space. Finally, \citet{2020fujimoto} found that the \cii emission extends up to 10 kpc, with an effective radius exceeding the rest-frame UV effective radius by a factor of 2, being thus identified as one of the individual detections of \cii haloes.

In summary, the galaxy is a typical star-forming galaxy in the first billion years of the Universe with a complex
morpho-kinematical structure and \cii extended emission. CRISTAL-05 is an ideal target to study mechanisms leading to \cii emission at large distances. The full description of the sample selection, science goals, and data products of the CRISTAL survey will be described in Herrera-Camus et al. (in prep).

\subsection{ALMA observations}

We combined the available  Band 7 ALMA observations of CRISTAL-05: ALPINE survey in configuration C43-2 (PI: Olivier Le Fèvre, ID: 2017.1.00428.L), CRISTAL pilot program in configuration C43-5 (PI: Manuel Aravena, ID: 2018.1.01359.S) and CRISTAL observations in the configurations C43-2 and C43-5 (PI: Rodrigo Herrera-Camus, ID:2021.1.00280.L). The observations have total on-source times of 22, 136, 92, and 24 minutes respectively. All observations aimed at detecting the \cii emission line ($\rm\nu_{obs} = 290.6$ GHz) and FIR dust continuum in four spectral windows. The different programs use slightly different spectral setups and array configurations but with the frequency range for the \cii line emission well covered. In all cases, four spectral windows were used with an individual usable bandwidth of 1875 MHz.

% alpine 1360 S manual  8164.800  CRISTAL:  5503.680   1451.520 

%%% Edited by MA

\subsubsection{Data reduction and JvM correction}

Data calibration and a combination of the different observations were performed using the CASA software \citep{2022casa}. We processed the datasets from the different programs using the corresponding pipeline versions: 5.6.1 for the ALPINE program, 5.6.1 for the CRISTAL pilot program, and finally, version 6.5.2 for the CRISTAL program. No extra manual flagging was needed beyond what was already identified by the observatory and automatically flagged by each pipeline. We combined the calibrated datasets into a single measurement set (ms) used to create the images and data cubes, using the task {\sc concat}. This task and the following procedure are performed using version 6.5.2 of CASA software. We identified the emission lines and continuum emission in the initial multi-frequency synthesis (mfs) images and data cubes of all the available spectral windows (SPWs) to then create the line-free continuum images and subtract the continuum emission using \textsc{uvcontsub} with {\sc fitorder = 0}.

We created the final product using task \textsc{tclean} and the auto-multithresh algorithm. This mode automatically masks regions based on the signal-to-noise of the emission in the image. How such regions are created depends on the following parameters: {\sc Sidelobethreshold = 2}, {\sc Noisethreshold = 4.5}, {\sc Lownoisethreshold = 2} and {\sc Minbeamfrac = 0.0}. This value choice is intended to match similar results when using manual cleaning. In all the cases, the cleaning was performed down to $1\sigma$ by selecting {\sc nsigma=1} where $\sigma$ is estimated automatically by \textsc{tclean} using robust statistics ($\sigma=1.4826\times \rm{MAD}$, with MAD being the median absolute deviation). We mainly worked on the products created using a Briggs weighting scheme \citep{1995briggs} with {\sc robust = 2}, but some auxiliary inspections were carried out on the products with {\sc robust = 0.5} (Section~\ref{sec:gasmotion}).  All the products have a pixel size of 0.0245\arcsec pix$^{-1}$, and the reference frequency of the \cii emission is the same as defined to the source in the ALPINE catalog, i.e., $\nu_{\cii} = 290.541$ GHz \citep{2020lefevre}.

The primary motivation for cleaning down to such a low threshold of {\sc nsigma=1} was to minimize the "JvM effect" \citep{JvM1995,Czekala2021}. This effect occurs when combining multiple array configurations of an interferometer, which can produce a non-ideal baseline distribution and a synthesized beam that will depart from an ideal Gaussian beam. In these cases, the {\sc TCLEAN} algorithm can produce final images that mix dirty and clean beam units and incorrect recovered fluxes. To correct for the JvM effect, we applied in all the products the correction suggested by \citet{Czekala2021} that compares the volumes of the clean and dirty beams to determine an $\epsilon$ correction parameter to apply to the residuals. In the case of the data cubes, the correction was estimated at the channel closer to the peak of the \cii line and applied to all channels.  The correction factors for all the products we have used in this article lie within the range of $\epsilon = 0.3 - 0.4$. 
To assess the influence of this effect on our measurements, the total flux density of CRISTAL-05 is approximately 2.5 $\times$ larger in the non-JvM-corrected cube compared to the corrected one when extracted within a circular aperture of 1.5\arcsec centered on the source. 
%As an identicator of how this effect impacts our measurements, the total flux density of \target is 160\% overestimated when calculating in a circular aperture of 1.5\arcsec centered in the source. 
A more detailed description of the data reduction and JvM correction is covered in Gonzalez-Lopez et al, in prep. We summarize the noise level and the synthesized beam sizes of the products used along the following sections in Table~\ref{tab:products}. 

Finally, one downside of using this correction is that it also affects the background of the images and cubes. The root mean square (rms) of the JvM-corrected background is reduced by a factor of $\epsilon$, no longer accurately reflecting the sensitivity for point-like sources. Consequently, for visualization purposes, we display the non-JvM-corrected products with contour levels representing the true sensitivity to point-like sources. While some images may exhibit emissions overestimated due to the JvM effect, all corresponding analyses were conducted using the JvM-corrected products.

\begin{table}
\caption{Observational and physical properties of CRISTAL-05}
\centering          
\begin{tabular}{l c }     % 7 columns 
\hline\hline       
                      % To combine 4 columns into a single one 
Property & Value \\ 
\hline            
   \\
   Center                                         & +10:00:29.86, +02:13:02.19    \\  
   $z_{\mathrm{[CII]}}$$^{a}$                     &  5.5420 $\pm$ 0.0005      \\ 
   SFR$_{\rm UV}$ (M$_{\odot}$ yr$^{-1}$)         &   22 $\pm$ 1                    \\ 
 
   SFR$_{\rm IR}$$^{b}$ (M$_{\odot}$ yr$^{-1}$)   &  1.38 $\pm$ 0.09   \\ 
   Stellar mass$^{b}$ ($10^{10}$ M$_{\odot}$)     &   1.4 $\pm$ 0.6               \\ 
    
   \\
   S/N$^{c}$ (\cii)                                     & 14                   \\ 
   
   \cii line flux (Jy km s$^{-1}$)                & 1.00 $\pm$ 0.07                \\  
   FWHM$_{\rm [CII]}$ (km s$^{-1}$)               & 238  $\pm$ 25                   \\ 
   S$_{\rm 158\mu m}$$^{b}$        (mJy)                &  0.14 $\pm$ 0.04          \\ 
   L$_{\rm [CII]}$ ($10^{8}$ L$_{\odot}$)         & 9.3  $\pm$ 0.7                  \\ 
   L$_{\rm IR}$$^{b}$  ($10^{11}$ L$_{\odot}$)           &  1.60  $\pm$ 0.08                \\ 
   \\ %3.6 $\pm$ 0.5

  % &  &  \\ 
   %&  &  \\ 
   %&  &  \\ 

\hline  
\hline

\end{tabular}
\flushleft{{\bf Notes:} Values derived in this work were obtained in a 1.5\arcsec-radius aperture. $^{a}$ \citet{2020bethermin}, $^{b}$Measured by \citet{2023Mitsuhashi}, $^{c}$Integrated S/N. } 
\label{tab:properties-central}  
\end{table}

\subsection{HST observations}

We use the available HST/WFC3 NIR imaging of CRISTAL-05 as part of the Cosmic Assembly Near-IR Deep Extragalactic Legacy Survey (CANDELS, PI: Sandra Faber, ID: 12440). Observations were performed in 2015 in 1 orbit per filter for the F105W (1.05 $\mu$m), F125W (1.25 $\mu$m), F140W (1.4 $\mu$m), and F160w (1.54 $\mu$m) filters. For a galaxy at z $\sim$ 5.5, these filters correspond to the rest-frame UV emission centered at 1605 \AA, 1911 \AA, 2150 \AA\ts and 2450 \AA, respectively. Raw images were re-calibrated using the latest versions of the {\sc grizli} pipeline and the astrometry was corrected using PAN-STARSS and GAIA stars, leading to an uncertainty of about 0.1\arcsec. All images have a point spread function (PSF) with a full-width half maximum of FWHM = 0.2\arcsec and a pixel scale of 0.06 \arcsec pixel$^{-1}$. The unobscured star formation rate SFR$\rm_{UV}$ is derived by
 the luminosity measured in the HST/F105W filter, using the calibration provided by \citet{2012kennicutt}, and assuming a Kroupa IMF \citep{2003kroupa}.

\subsection{JWST observations}

JWST observed CRISTAL-05 as part of the PRIMER program \citep[GO 1837, ][]{2021dunlop} using the Mid-Infrared Instrument (MIRI) camera, and it was observed for 29.6 minutes. The filter used was F770W, corresponding to the rest-frame near-infrared emission at 1.2 $\mu$m. The image has a point spread function (PSF) with a full-width half maximum of FWHM = 0.269\arcsec and a pixel scale of 0.049\arcsec  pixel$^{-1}$. The data were calibrated using the latest version of the pipeline and astrometrically corrected using both GAIA stars and the sources detected by HST in the same field. 

\section{Results}

\begin{figure}
    \centering
    \includegraphics[scale= 0.5]{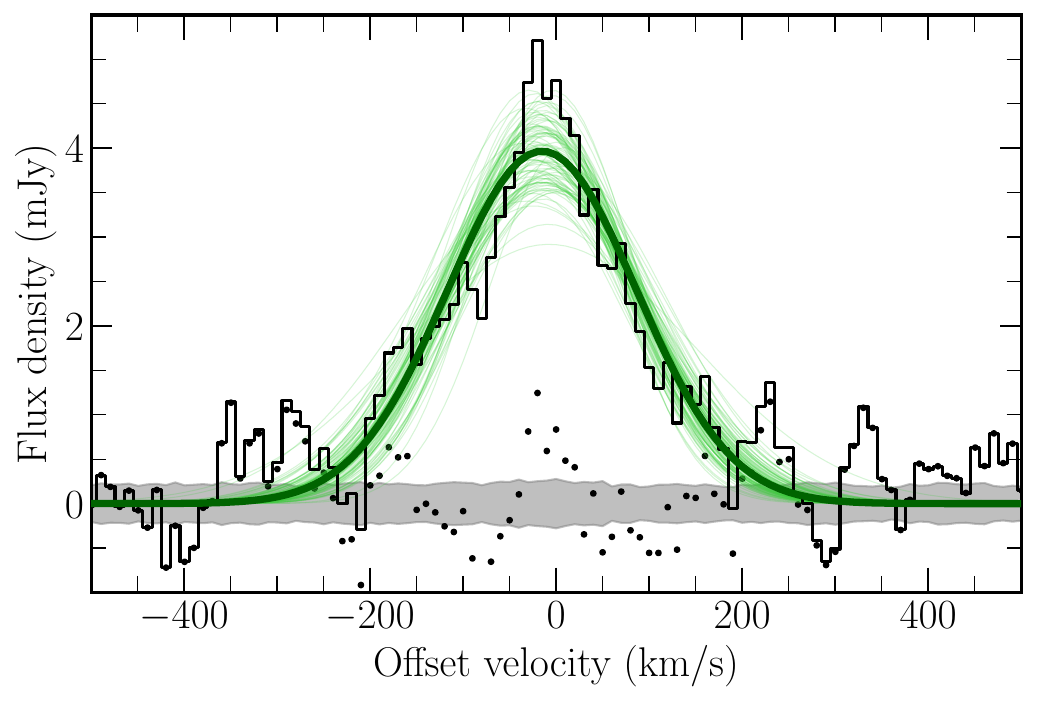}
    \includegraphics[scale= 0.5]{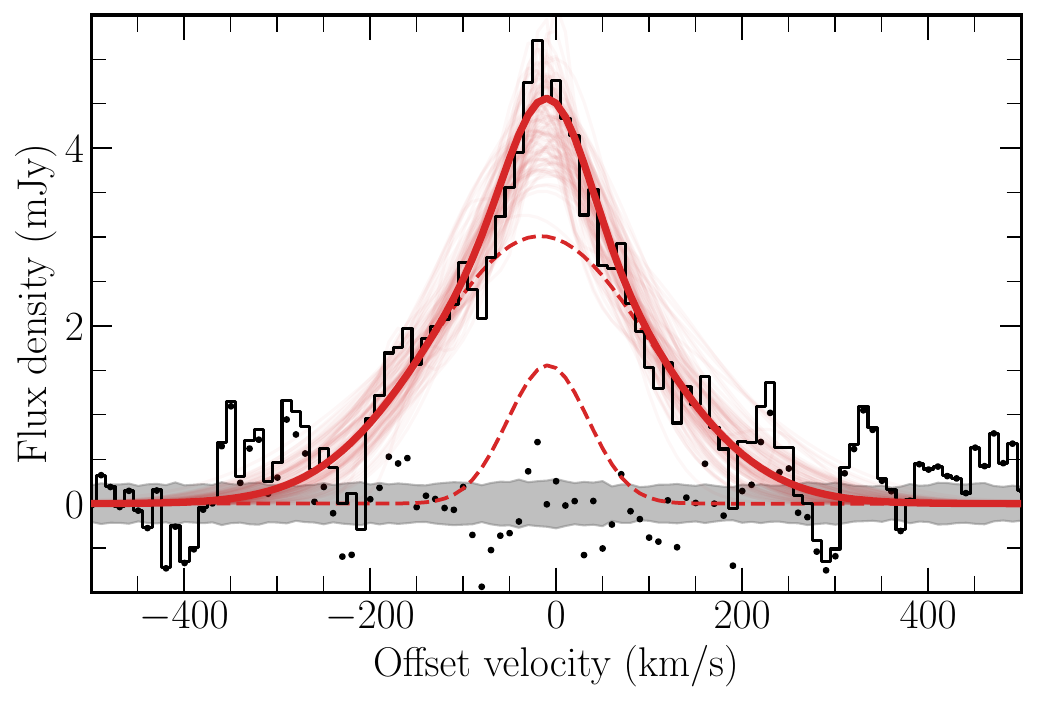}
   \caption{JvM-corrected \cii spectrum of CRISTAL-05, which is extracted within a circular aperture with a radius of 1.5\arcsec, and centered in the \cii observed frequency $\rm\nu_{obs}$ = 292.48 GHz. We display the continuum-subtracted spectrum from the JvM-corrected data cube, with a channel resolution of 10 km~s$^{-1}$  (black solid lines in both panels). (a) Top: The green solid line corresponds to the 1-D Gaussian fitting, and the residuals are plotted as black dots. (b) The red solid line corresponds to the fitting of the spectrum with two Gaussian components. As in the panel above, the residuals are plotted as black dots.}
    \label{fig:spectrum-cii}
\end{figure}

\begin{figure}%[h]
    \centering
     \includegraphics[width=0.4\textwidth]{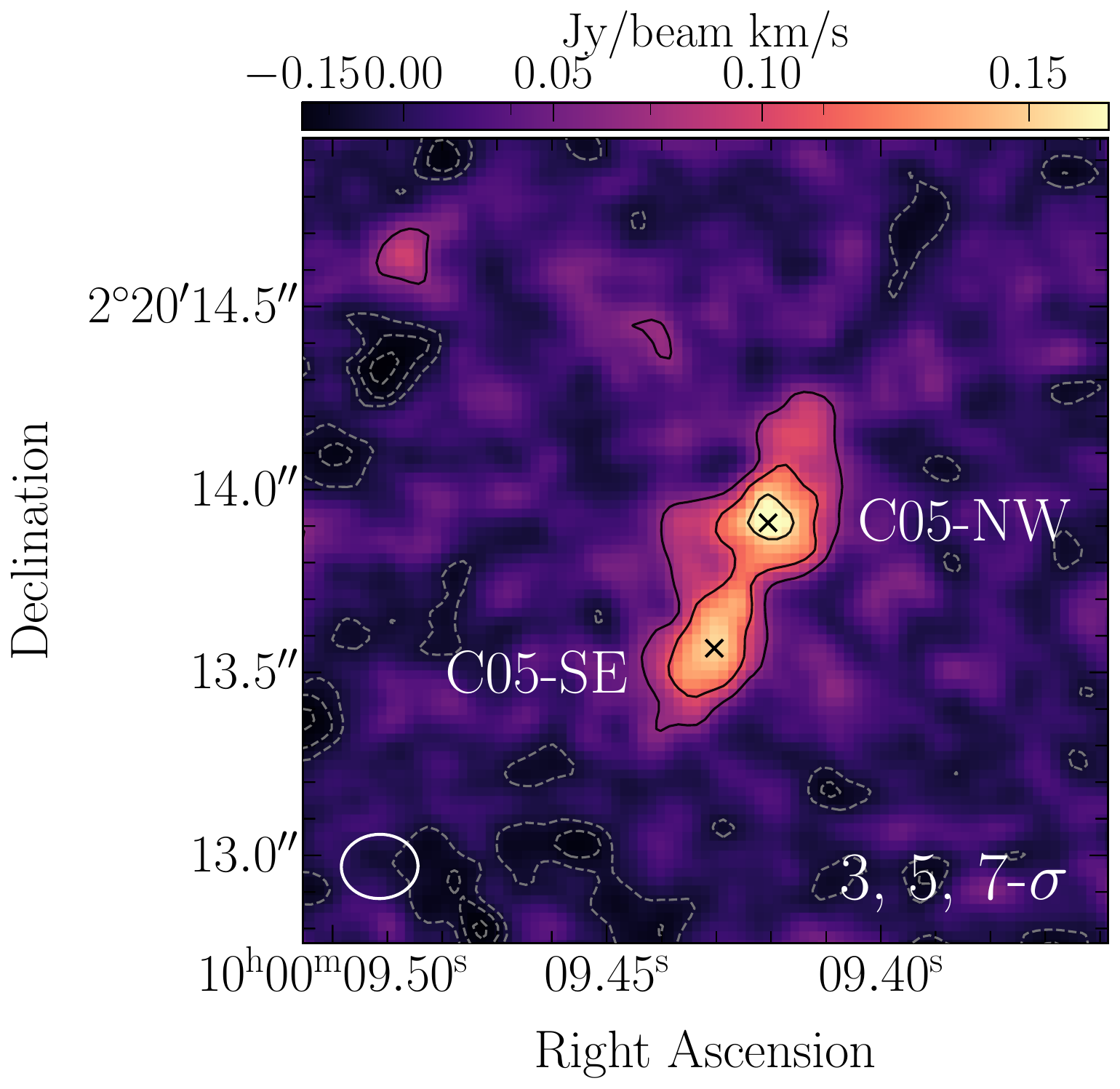}
    \caption{ ALMA \cii moment-0 map (non-JvM corrected). The overlaid black contours correspond to the 3, 5, 7-$\sigma_{\cii}$ levels, where $\sigma_{\cii} = $ 0.02 Jy/beam km s$^{-1}$ is the rms noise level. The white ellipse indicates the beam size of $0.21\arcsec\times 0.18\arcsec$. }
    \label{fig:cii-briggs}
\end{figure}

\subsection{\cii line and dust emission}\label{sec:cii-line}

We measured the spatially integrated \cii line profile using a 1.5\arcsec-aperture centered at the emission centroid (shown in Table~\ref{tab:properties-central}). \cii is detected and best fit by a single-gaussian distribution with FWHM = 238 $\pm$ 25 km s$^{-1}$ and line flux of 1.00 $\pm$ 0.07 Jy km s$^{-1}$, corresponding to a signal-to-noise of S/N = 14. This is consistent with the total flux measured by the ALPINE survey \citep{2020lefevre}, and about 45\% higher than the total flux of observations of \citet{2015capak}. The JvM-corrected \cii spectrum is shown in Figure~\ref{fig:spectrum-cii} top panel (black solid line), and the best-fit single Gaussian as the green line. The moderate goodness of the fit ($\rm\chi^{2}$ = 6.5) is explained by the heavy-tail shape of the \cii distribution, with a blue wing around v $<$ -100  \kms.  A combination of two Gaussian components better fits the wing and the peak of the emission ($\rm\chi^{2}$ = 5.2), as shown in the bottom panel of Figure~\ref{fig:spectrum-cii}. 

The dust emission was measured by \citet{2023Mitsuhashi}, yielding a flux density of S$_{158\mu m}$ = 0.118 $\pm$ 0.021 mJy. This is a detection with a significance of S/N = 6.7. The global properties of the CRISTAL-05, such as spatial centroid, line width, and integrated flux presented is summarized in  Table~\ref{tab:properties-central}.

\subsection{Gas motion}
\label{sec:gasmotion}

\begin{figure*}
    \centering
    \includegraphics[width=\textwidth]{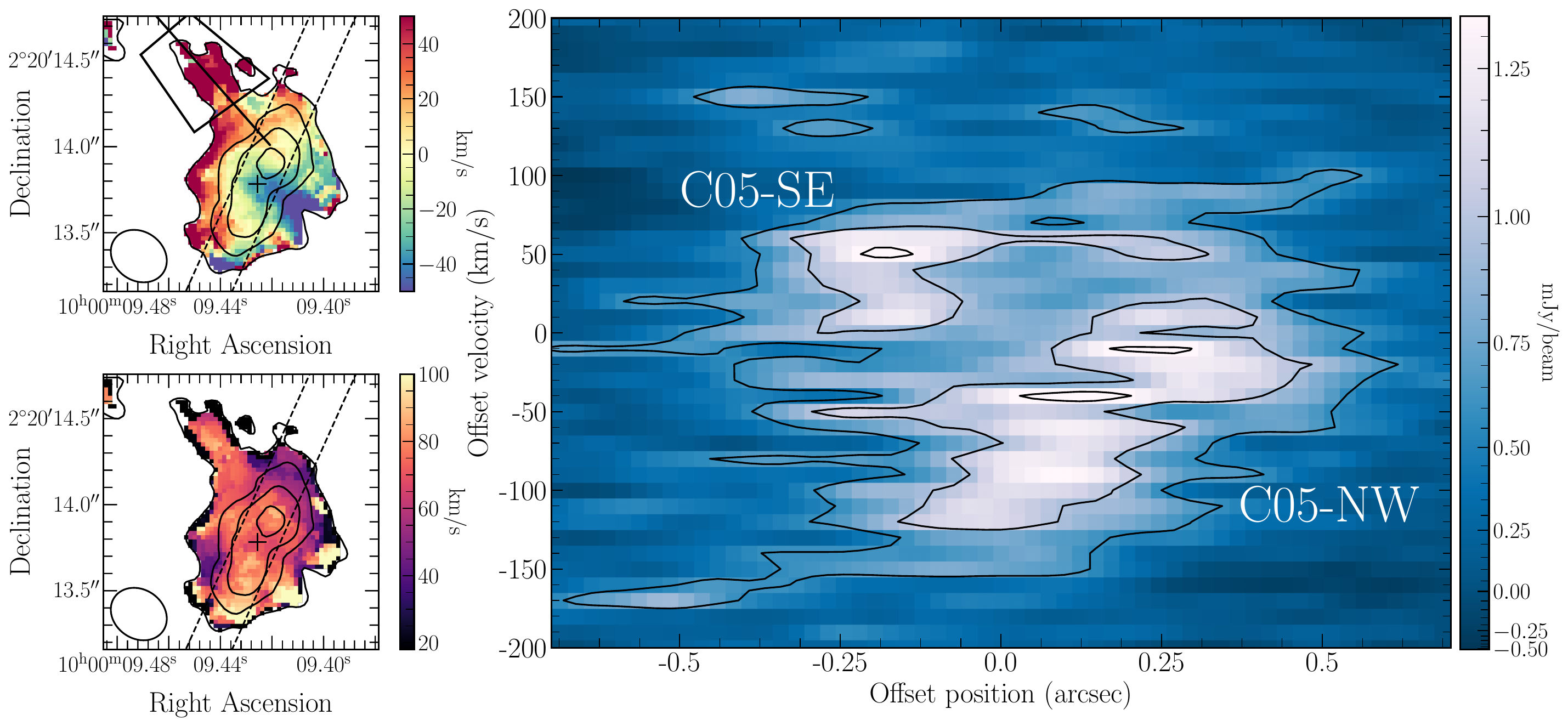}
    \caption{Kinematics of the \cii emission for CRISTAL-05.  (a) Top and (b) bottom left panel:  it is respectively shown the velocity (moment-1) and dispersion (moment-2), with respect to the observed frequency of \cii at redshift z = 5.54. The black contours correspond to the 3-, 6-, 9-, 12-$\sigma$ levels of the \cii moment-0 map (non-JvM corrected). The black dashed line shows the slit in which we created the position-velocity diagram in the right panel, with the center of the slit as a black cross. The solid black line (black polygon) represents the slit used to generate the pv diagram (spectrum) of Figure~\ref{fig:arm-pv-spec}. (c) Right panel: Position-velocity diagram (non-JvM corrected) along the major axis of CRISTAL-05 with a width of 0.245\arcsec. The black contours represent the 3-, 5-, 7-$\sigma_{pv}$.}
    \label{fig:moment-1-2}
\end{figure*}

\begin{figure}[h!]
    \centering
    \includegraphics[width=0.43\textwidth]{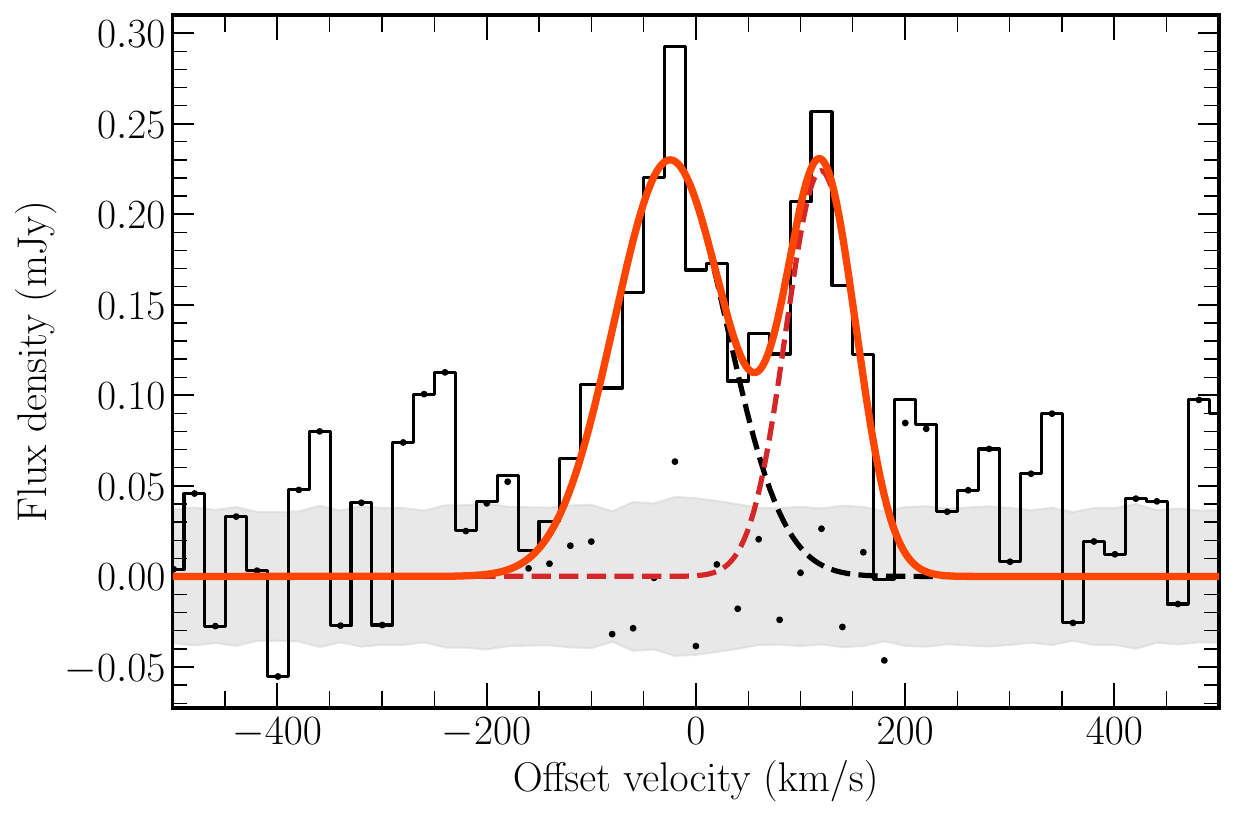}
    \includegraphics[width=0.5\textwidth]{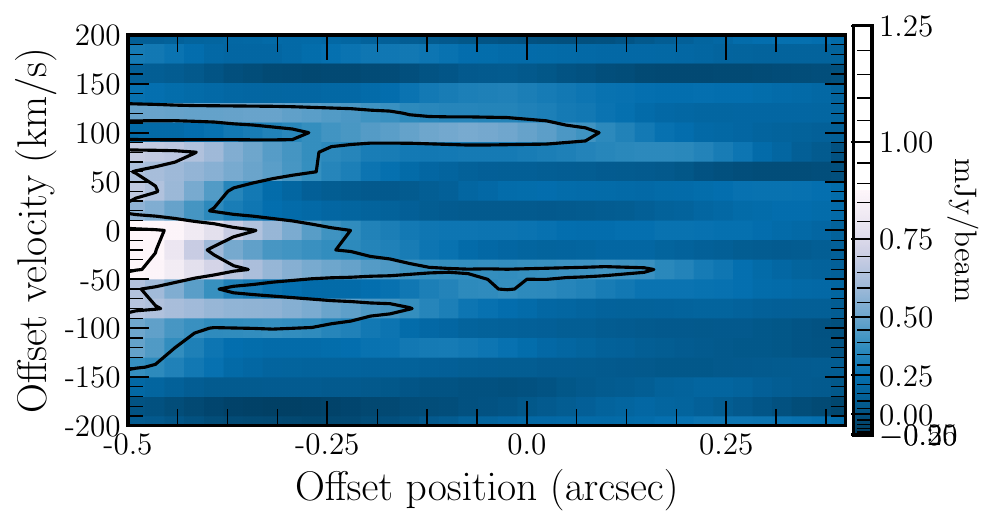}
    \caption{Kinematical properties of the northeast emission of CRISTAL-05. (a) Top panel: \cii spectrum within the black polygon in the top
left panel of Figure~\ref{fig:moment-1-2}.  The orange solid line corresponds to the spectrum fitting with two Gaussian components, and the residuals are plotted as black dots. (b) Position-velocity diagram along the slit, shown as the black solid line in Figure~\ref{fig:moment-1-2}, with a width of 10 pixels.}
    \label{fig:arm-pv-spec}
\end{figure}

We show in Figure~\ref{fig:cii-briggs} the non-JvM-corrected \cii moment-0 map, derived by collapsing the cube over the full-width tenth maximum of the \cii line. We aim to resolve the internal structure of CRISTAL-05, therefore we display the cube generated by a Briggs weighting ({\sc robust = 0.5}), yielding a final beam size of 0.21\arcsec x 0.18\arcsec (1.2 x 1.0 kpc). Previous ALMA observations \citep{2015capak,2020lefevre} yielded a significant yet mostly unresolved detection (0.9\arcsec $\sim$ 5 kpc) of the \cii emission line in CRISTAL-05. The current spatial distribution of the \cii emission differs significantly from these former observations, as CRISTAL-05 now exhibits a double-peaked emission, originating from two components aligned at 25 degrees clockwise from the north axis. The locations of these peaks are denoted by black crosses in Figure~\ref{fig:cii-briggs}. In the following sections, we refer to the northern emission distribution as the C05-NW component and the southern emission as the C05-SE component. The C05-NW peak is slightly brighter than the southern one, and their separation is approximately 0.4\arcsec ($\sim$ 2.4 kpc), at least two times smaller than the resolution of the previous observation.

The double-peak shape, revealed by our deeper and more resolved observations, suggests that we are observing a multicomponent system, potentially arising from a close approach of two galaxies or the presence of internal clumps within a single galaxy. We investigate these scenarios by examining the kinematic properties of the gas. High-resolution observations are susceptible to surface brightness dimming. This dimming effect can impact the detection of faint features, such as diffuse gas. To mitigate this possibility, we employ data products generated through natural weighting in the next analyses. This corresponds to a typical beam size of 0.3\arcsec, equivalent to 1.8 kpc at the redshift of CRISTAL-05.

We created intensity-weighted velocity maps (moment-1) and intensity-weighted dispersion velocity maps (moment-2) of the \cii emission, as depicted in the top left and bottom panels of Figure~\ref{fig:moment-1-2}. These moments were computed along the channels within the velocity range of [-280,+280] km s$^{-1}$, encompassing all the emission of the \cii line (see Figure~\ref{fig:spectrum-cii}). The averaging only includes pixels in a given channel with an intensity greater than 3 $\times$ the rms of the channel. Additionally, we generate a position-velocity diagram to illustrate how gas motions vary with the position along a predefined slit, as shown in the right panel of Figure~\ref{fig:moment-1-2}. The slit is positioned along the major axis of the CRISTAL-05 system  (the region between the dashed black lines in the left panels of Figure~\ref{fig:moment-1-2}), with a width of 10 pixels, approximately 0.245\arcsec.

Examining the moment-1 emission map (left top panel in Figure~\ref{fig:moment-1-2}), it is evident that the line-of-sight velocity of the gaseous component is not uniformly distributed across the main system, primarily ranging from -50 to 50 km/s. Along the minor axis of the system, the gas component on the east side of the galaxy appears redshifted from the central velocity of the line ($>$ 20 \kms), while the west side presents negative velocities. Similarly, focusing on the central region encompassing both sources (within the black contour exceeding 6-$\sigma$), a velocity gradient is observed along the major axis for each source. The moment-2 map (bottom left panel of Figure~\ref{fig:moment-1-2}) reveals a relatively uniform dispersion velocity but with a notably high average value of $79\pm 21$ \kms.

An inspection of the position-velocity diagram (right panel of Figure~\ref{fig:moment-1-2}) reveals that CRISTAL-05 exhibits a complex kinematic structure comprising at least two components, surrounded by disturbed carbon-rich gas. The overall \cii emission spreads over a projected angular size of 1\arcsec and, in contrast to the moment-1 map, spans a total line-of-sight velocity from -150 to 100 \kms. We can easily pinpoint the two components C05-NW and C05-SE. The C05-NW component extends over 0.7\arcsec, equivalent to a physical size of approximately 4 kpc, and over a velocity range from nearly -150 km s$^{-1}$ to 30 km s$^{-1}$. This component is the most extended and broad, suggesting that this is the dominant object in the system. It also exhibits a disturbed shape, likely caused by the proximity to C05-SE. Interestingly, the peak of the emission of C05-NW rises in velocity at different positions along the slit. Such velocity gradient is commonly interpreted as a signature of rotation or, alternatively, the presence of two unresolved interacting objects \citep{2020forster}. The nature of C05-NW will be further investigated in the next sections. The second subdominant component, C05-SE, is more compact and narrower in the velocity axis, extending for 0.275\arcsec with $\Delta$v $\sim$ 60 \kms. A velocity gradient is not noted in this component, which means that this feature in the south part of the moment-1 map is just a projection effect. 

The fact that the two components considerably overlap along the line-of-sight spatially (from -0.25 to 0\arcsec), and present distinct velocity distributions explains why we note a narrow velocity range in the moment-1 map (-50 to 50 km~s$^{-1}$), but a high average dispersion velocity in the moment-2 map. C05-NW dominates the negative velocity range, while C05-SE shows only positive velocities such that the intensity-weighted velocity averages to values between those of the components and this leads to smaller velocity ranges in the moment-1 map. For the moment-2 map, the different velocity distributions result in broader intensity-weighted line widths along the line of sight across the system.

Lastly, we investigate the kinematic properties of the northeast emission of CRISTAL-05, which encompass the black polygon in the top left panel of Figure~\ref{fig:moment-1-2}. In the moment-1 map, the area exhibits mostly emission redshifted to velocities of $>$ 50 km s$^{-1}$.  Figure~\ref{fig:arm-pv-spec}  shows the spectra (top) calculated within the area of the polygon and the pv diagram (bottom) across the slit as a solid black line in the top left panel of Figure~\ref{fig:moment-1-2}. The black contours pf the pv diagram represent the 3-, 5-, 7-$\sigma_{pv}$ significance levels. The \cii spectrum shows a double peak emission centered at -25 km s$^{-1}$ and +120 km s$^{-1}$, both detectable in the top and bottom panels.  A Gaussian fitting reveals that the blue-shifted and redshift emission have flux densities of 3.1 $\pm$ 0.5 mJy km s$^{-1}$ and 1.88 $\pm$ 0.004 mJy km s$^{-1}$,  which corresponds to a significance level of  6-$\sigma$ and 4-$\sigma$, respectively. This feature is an indication of the disturbed nature of the system. 

%In brief, the morpho-kinematical distribution of the \cii emission reveals that \target is a system containing at least two gaseous components. This associated with the disk shape of the moment-0 map (natural weighting) can be explained by the inner cores of gaseous internal structures of a clumpy single galaxy, or by a galaxy merger.  However, the non-uniform line-of-sight velocity distribution of the moment-1 map and the overall high dispersion velocity of the moment-2 rule out the possibility that this is a single system. The position-velocity diagram clarifies that the two components have two distinct velocity profiles, although the sources spatially overlap. The emission distributions are substantially disturbed, and we can also observe disturbed gas surrounding the system, favoring that \target has signs of being a merging galaxy. 

\subsection{Multicomponent morphologies} \label{sec:morphology}

\begin{figure*}
    \centering
    \makebox[\textwidth][c]{\includegraphics[width=1\textwidth]{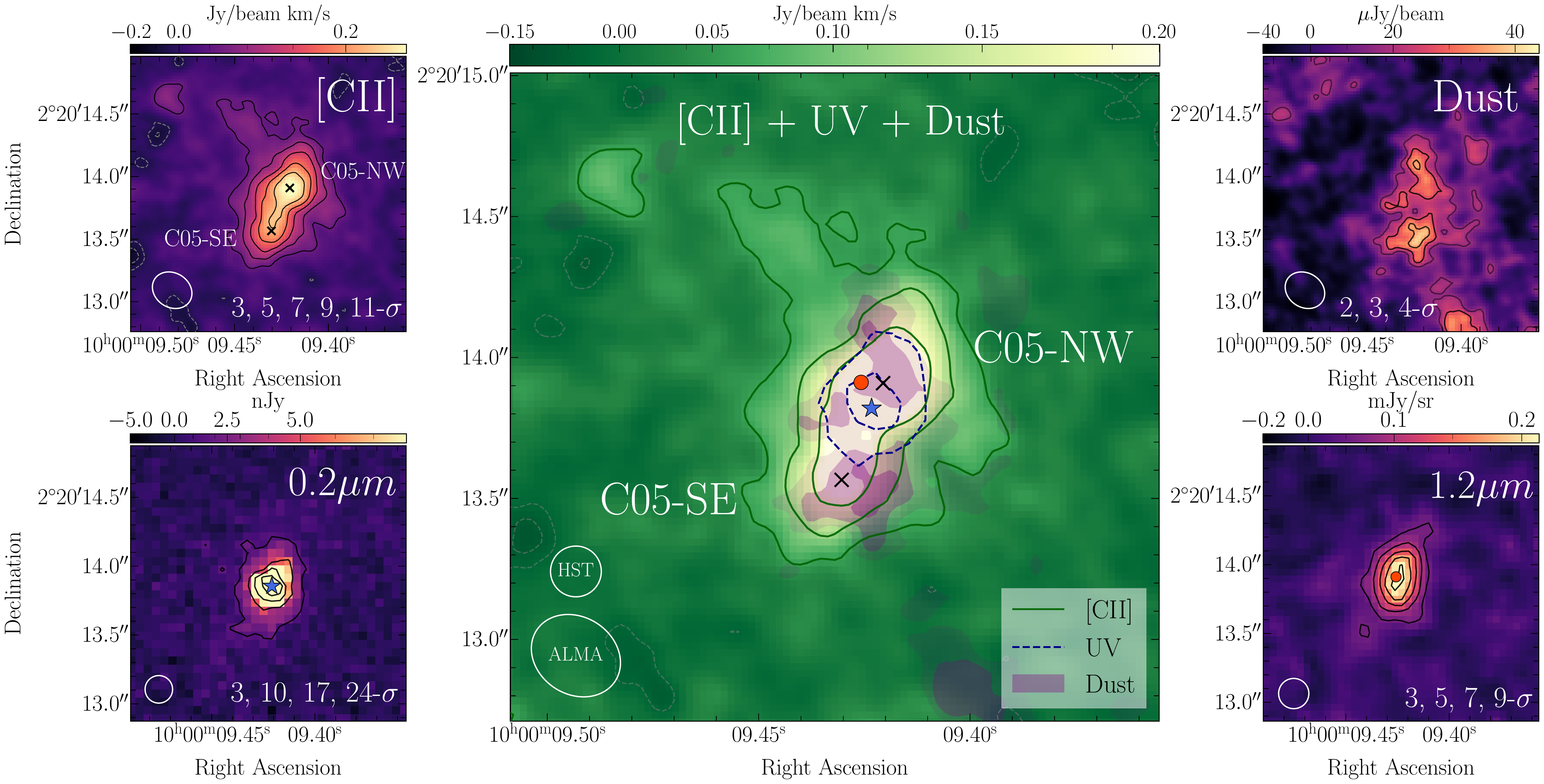}}

    \caption{Multi-wavelength cutout postage stamps toward CRISTAL-05.  {(a) Left top panel:} ALMA \cii moment-0 map (non-JvM corrected). The overlaid black contours correspond to the 3, 5, 7, 9, and 11-$\sigma_{\cii}$ levels. The white ellipse indicates the beam size of $0.33\arcsec\times 0.27\arcsec$. {(b) Left bottom panel:} HST F140W rest-frame UV (0.2$\mu$m) continuum image. The overlaid contours correspond to the  3, 10, 17, and 24-$\sigma_{UV}$ levels. The white circle indicates the 0.2\arcsec point spread function. The peak emission is marked as a blue star. {(c) Right top panel:} non-JvM-corrected ALMA rest-frame 160$\mu$m dust continuum image. The overlaid contours correspond to the  2, 3, 4-$\sigma_{160}$. The white ellipse indicates the beam size of $0.33\arcsec\times 0.27\arcsec$, the same as the \cii moment-0 resolution.  {(d) Right lower panel:} JWST F770W (1.2$\mu$m) rest-frame near-infrared image. The black contours correspond to 3, 5, 7, 9-$\sigma_{1.2}$. The peak emission is marked as a red circle. (e) Central panel: non-JvM-corrected ALMA \cii moment-0 map in the background, similar to the left top panel. We display the \cii peak emission of  C05-NW and C05-SE components as black crosses (based on Figure~\ref{fig:cii-briggs}). The overlaid green contours correspond to the 3, 6,  and 9-$\sigma_{\cii}$ levels. The blue dashed lines represent the HST/F140W map, for the 7- and 20-$\sigma$ significance levels. The filled areas correspond to the 160$\mu$m dust continuum emission for the 2-, 3- and 4-$\sigma$ significance levels.}

                    %{ \color{darkmagenta} aumentar o label, uv: adicionar psf e label dos levels  } 

    \label{fig:moments}
\end{figure*}

In this section, we characterize how star formation (as traced by FIR and rest-frame UV), stellar content (rest-frame optical), and cold gas are distributed across the CRISTAL-05 system. Figure~\ref{fig:moments} shows the \cii moment-0 map collapsed over the full-width tenth maximum of the \cii line (top left panel), the rest-frame UV at 0.2$\mu$m (traced by the HST F140W-band image; bottom left), the rest-frame FIR dust continuum emission at 160$\mu$m (traced by the observed 850$\mu$m continuum; top right), and the rest-frame near-infrared at 1.2$\mu$m (traced by the JWST F770W image; bottom right). The central main panel gathers the information on all the different wavelengths.

The \cii emission is spatially extended across $\sim10$ resolution elements (top left panel). Although this image has a lower resolution than the Briggs image ({\sc robust = 0.5}, Figure~\ref{fig:cii-briggs}), we can still recover the double peak shape, with the peak emission of the whole system coinciding with the location of component C05-NW. The major axis of the source extends over 1.2\arcsec ($\sim$ 7 kpc), which corresponds to be resolved in 8 resolution elements. The emission is elongated from the southeast to the northwest, and now it shows an asymmetric, disturbed shape. As we noted in the previous section, we find emission with significance greater than 3$\sigma$ extending from the main source towards the northeast, and there is also an emission ($>3\sigma$) distant 1\arcsec in the northeast of the galaxy.

Contrary to the \cii emission, the rest-frame UV image (bottom left panel in Figure~\ref{fig:moments}) shows a compact single source, slightly elongated in the northwest-southeast direction.  The peak of the rest-UV emission, shown as a blue star in the left-bottom and central panels of Figure~\ref{fig:moments}, is closer to the peak of C05-NW  (0.1\arcsec $\sim$ 0.6 kpc, upper black cross in the central panel) than the C05-SE component (0.3\arcsec $\sim$ 1.8 kpc). However, some rest-UV emission extends in the region between both \cii source peaks (blue dashed line in the central panel of Figure~\ref{fig:moments})

Also shown in Figure~\ref{fig:moments} is the JWST $1.2\mu$m image, which traces the rest-frame NIR close to the peak of the stellar emission at $z=5.5$ (rest-frame $1.8\mu$m). This stellar emission is clearly detected and, as with the rest-frame UV, is found to be compact, elongated along the direction of the \cii emission, and peak  (red circle in the central panel) located also close to the C05-NW component (0.07\arcsec $\sim$ 0.4 kpc).

Lastly, dust continuum emission is detected with a lower significance than \cii (peak S/N$\sim3\sigma$) throughout the system. Two components are detected and aligned in the north-south axis, similar to the \cii emission. As shown in the central panel of Figure~\ref{fig:moments}, the rest-UV emission peaks right in between the two dust components or clumps, which closely follow the location of the \cii C05-NW and C05-SE components. This suggests that dust obscuration can be one of the causes for the compactness of the rest-frame UV emission and/or that C05-SE is a dust-obscured source.  We also detect an emission greater than 3-$\sigma$   in the south part of the top right and central panels, with no counterpart of UV or \cii emissions, possibly from a source in a different redshift.

\section{Analysis}

\subsection{Radial profiles}\label{sec:radial-profile}

\begin{figure}
    \centering
    \includegraphics[scale = 0.4]{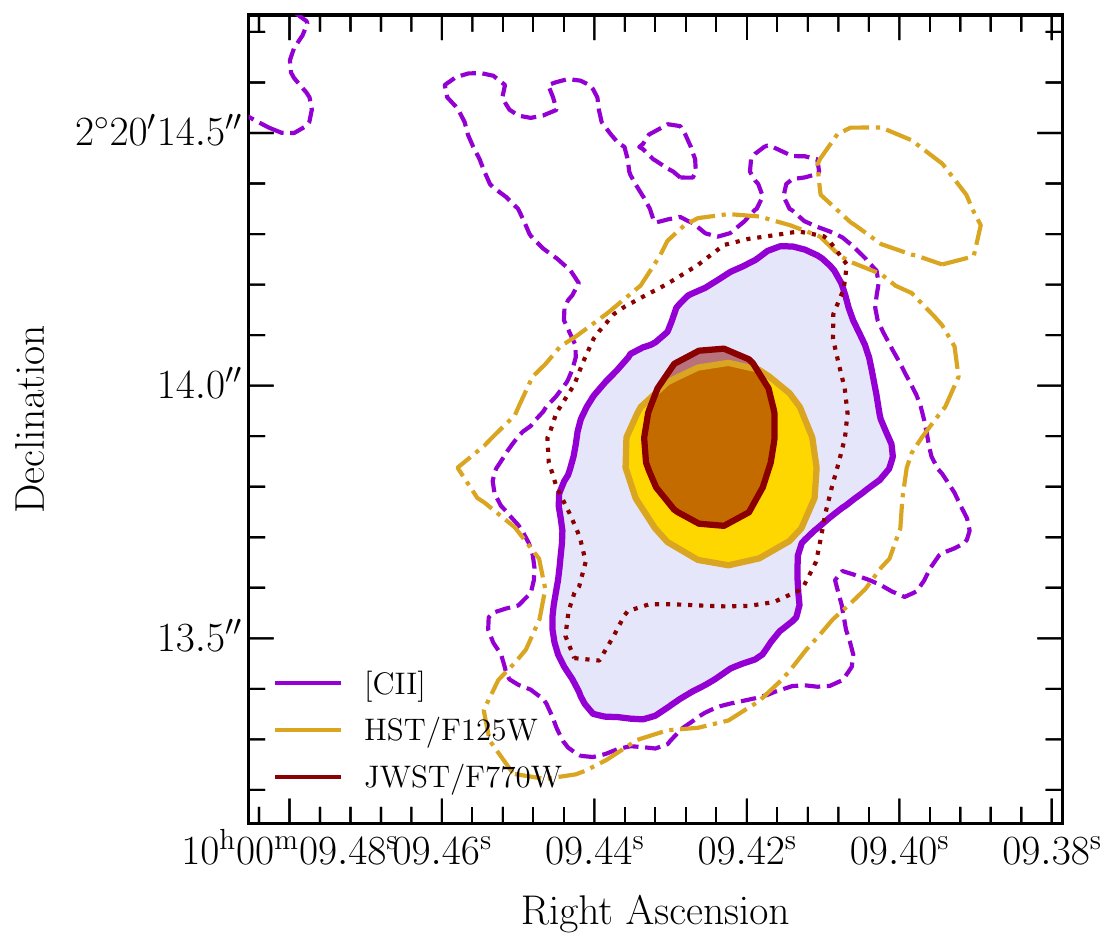}
    \includegraphics[scale = 0.4]{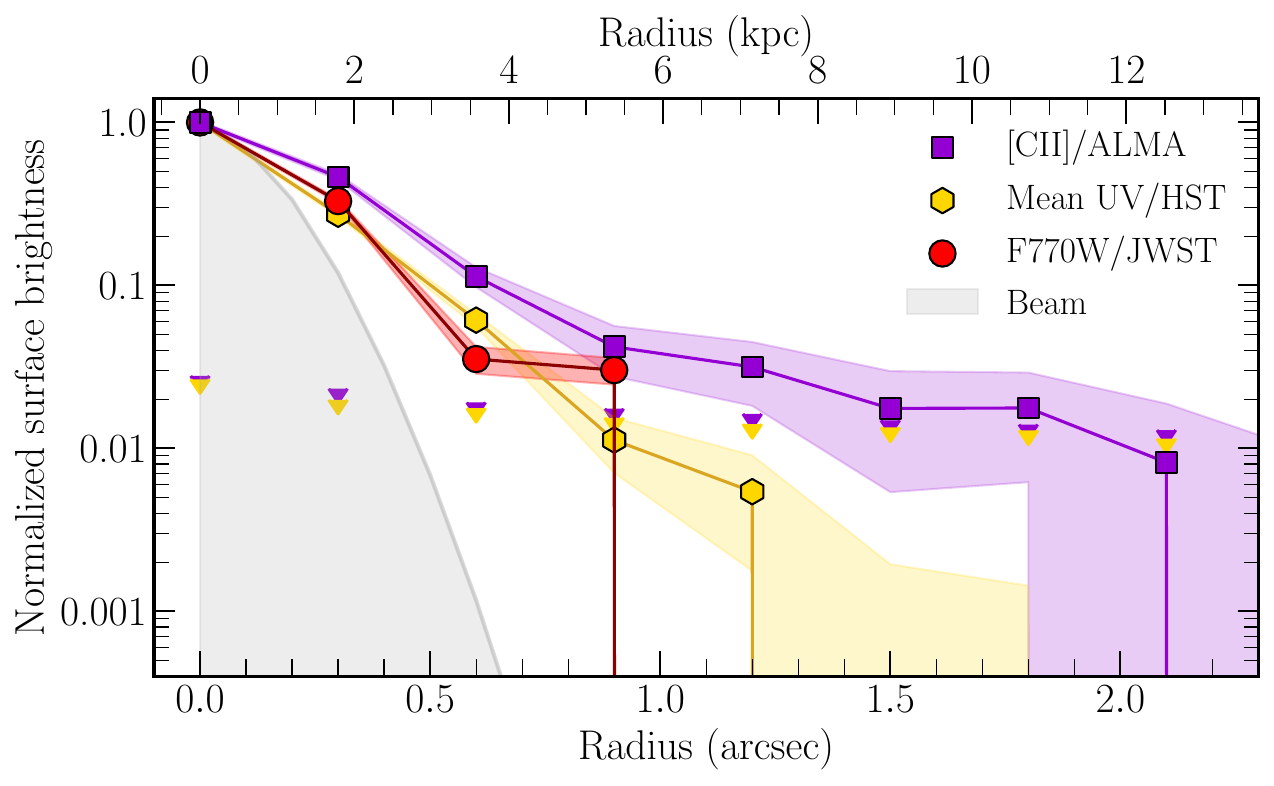}
    \caption{Comparison of the multi-wavelength radial profiles. 
    (a) Top: Resolution-matched (0.33\arcsec x 0.27\arcsec) moment-0 maps.  The  \cii, HST, and JWST are color-coded by purple, yellow, and dark red, respectively. The solid lines correspond to the significance levels, which contain half of the total flux density of the galaxy. The dashed, dotted-dashed, and dotted lines correspond to the 3$-\sigma$ edge of the source in \cii, rest-frame UV, and rest-frame NIR, respectively. (b) Bottom: Surface brightness profiles of \cii and (average F105W + F125W + F140W + F160W) UV emission, in 0.3\arcsec-width elliptical apertures all centered in the \cii centroid. The purple and yellow arrows show 1$\sigma$ upper limits.}
    \label{fig:radialprofile}
\end{figure}

CRISTAL-05 has compact star-forming and stellar components compared to the gaseous \cii emission. To quantify these extensions, we compare the effective radii and surface brightness radial profiles of all structural elements. Due to the proximity of the components and following \citet{2020fujimoto} analysis, we evaluate how the emission is distributed along the system considering it as a single source. The \cii,  rest-frame UV and near-infrared images have different resolutions, therefore we degraded the PSF of the HST and JWST maps to the ALMA  beam size, which is the lowest resolution among the data sets. It yields a final beam size of $0.33\times0.27\arcsec$ ($2.0\times1.6$ kpc$^2$).

We first compare the spatial distribution of the different components in the top panel of Figure~\ref{fig:radialprofile}. Highlighted as solid-filled regions, we show the areas containing half of the total flux density for the \cii (violet), HST/F125W (yellow), and MIRI/F770W (dark-red) images. The dashed (violet), dash-dotted (yellow), and dotted (dark red) lines represent the 3-$\sigma$ flux contours in each band. We find that the half-light \cii emission is more extended than the rest-frame UV and near-infrared emission traced by the HST and JWST bands, consistent with the results of \citet{2020fujimoto}. Based on visibility modeling, a S\'ersic  profile with an index of $n=1$ (Ikeda et al. in prep) results in a circularized effective radius of $r_{\cii}^{\rm cir} = 1.6 \pm 0.3$ kpc. Similar modeling of the rest-frame UV emission yields $r_{\rm HST/F160W}^{\rm cir} = 0.36 \pm 0.02$ kpc  \citep{2023Mitsuhashi}  for the HST/F160W, making the \cii emission $\sim4 \times$ larger than the rest-frame UV. Although the 3-$\sigma$ flux contours of the \cii and rest-UV extend to similar radii, where no detection of rest-NIR emission is observed, this can be just a direct consequence of the depth of the observations.

We also computed surface brightness radial profiles of the \cii,  UV, and NIR emissions as shown in Figure~\ref{fig:radialprofile} (bottom), respectively seen as violet squares, yellow hexagons, and red circles.  We calculate the flux in elliptical annuli, with 0.3\arcsec widths and the axial ratio of q = 0.6, based on the deconvolved axis of a spatial Gaussian fitting of the \cii emission in the moment-0 map. The profiles are a function of distance to the \cii centroid of the whole system, which is close to (but not exactly at) the peak of emission for the dominant C05-NW component, and less than 1 kpc ($\sim 0.1$\arcsec) from the centroid of the rest-UV and rest-1.2um emission. The surface brightness and uncertainties are normalized by the value of the first aperture of each band. 
The shaded area for each radial profile represents the 1-$\sigma$ uncertainty in these measurements. 
%For each annulus, this is assumed to be the standard deviation of the surface brightness of hundreds of apertures placed in the image with no significant emission.
We find that the \cii emission has a more gradual decline with radii compared to the steeper decrement of the emission in the rest frame UV and NIR bands. The \cii emission extends significantly up to 10 kpc, beyond where the emission is dominated by noise, and is undetected at further than 12 kpc. The radial profiles of both the HST and JWST bands are consistently similar out to 4 kpc, thereafter the JWST/F770W flux falls below the noise level. This noteworthy distinct extension confirms the measurements based on the visibility modeling of the emission distribution (Ikeda et al. in prep) and is consistent with recent findings of extended \cii emission in massive star-forming galaxies at these redshifts \citep{2020fujimoto,2021herrera-camus,2022akins}. 

Due to the moderate spatial S/N, we can not perform a direct comparison of the radial profile of the dust distribution. 
As an attempt, we convolve in the image plane all the images to a resolution of 0.5\arcsec to increase the significance of the emission, but the dust surface brightness profile is dominated by noise at r $>$ 3 kpc.  
Based on a visibility-modelling of the 0.5\arcsec-tapered data, \citet{2023Mitsuhashi} directly measured the extension of the dust continuum emission leading to an effective radius of r$_{eff} = 1.4 \pm 0.6$ kpc, similar to the \cii extension.  Therefore the dust-obscuration may play a role in the compact UV distribution observed.

\subsection{Clumpiness}\label{sec:clump-fellwalker}

\begin{figure*}
    \centering
    \includegraphics[width=1.\textwidth]{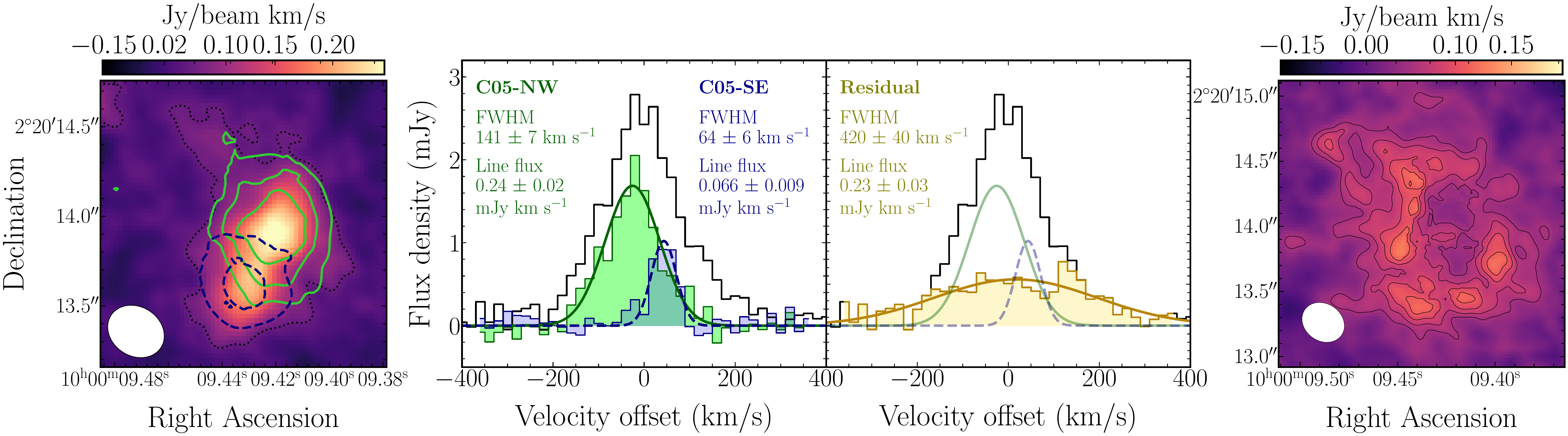}
    \caption{Morphologies and spectra of the individual galaxies and extended emission. [From left to right] (a) First panel: non-JvM-corrected \cii moment-0 map of CRISTAL-05, similar to the central panel of Figure~\ref{fig:moments}, with C05-NW and C05-SE maps overlaid as solid green and dashed navy lines, respectively (3-, 9-, 15-$\sigma_{rms}$). (b) Second panel: JvM-corrected spectra of CRISTAL-05 (black line), C05-NW (green line), and C05-SE (navy line), for apertures that enclose all the emissions of each component. The Gaussian distribution represents the best fit for the spectra of C05-NW and C05-SE, with a FWHM of 141 $\pm$ 7 and 64 $\pm$ 6 \kms, respectively. (c) Third panel: JvM-corrected spectra of CRISTAL-05 (black line) and the subtraction of components C05-NW and C05-SE (yellow line). The Gaussian distribution represents the best fit for the spectra of the residual, with a FWHM of 420 $\pm$ 40. (d) non-JvM-corrected \cii moment-0 map of CRISTAL-05, with the residual emission of the cube after subtracting components C05-NW and C05-SE (as 3-, 9-, 15-$\sigma_{rms}$)} %3,9,15
    \label{fig:residual-fellwalker}
\end{figure*}

The \cii moment-0 map and position-velocity diagram reveal that CRISTAL-05 is an interacting system, consisting of at least two galaxies projected along the same line-of-sight. Based on the visual inspection of the position-velocity diagram (Figure~\ref{fig:moment-1-2}), C05-NW shows hints of hosting a rotating gaseous disk. However, at this point, we cannot distinguish it from a scenario where C05-NW is, in fact, two components in a close approach that mimics the kinematics of a single rotating galaxy. In this section, we aim to determine how many components form CRISTAL-05.

We used the Fellwalker algorithm \citep{2005berry} to search for structures based on a gradient-tracing scheme. In brief, the algorithm walks around pixels in the 3D cube that exceeds a specific threshold (determined by the NOISE parameter). It searches for paths with the steepest ascents, i.e. it continually jumps from a given pixel to a neighboring pixel with the highest flux value heading to the peak flux value of the path. All paths reaching the same peak pixel are associated with the same clump. To define the starting point of each path, the algorithm considers only the section with an average gradient over four pixels greater than a threshold (determined by the FLATSLOPE parameter). Once a peak is found, the code expands its search to a broader neighborhood (defined by the MAXJUMP parameter) to check if there are pixels with higher values, avoiding noise spikes that may create local maxima. Adjacent peaks with pixel values below a certain threshold (defined by MINDIP) are thus considered part of the same clump.

We start the clump search with the hypothesis that only two or three clumps are present. All runs were performed in the version of the non-Jvm-corrected data cubes with a channel resolution of 10 km~s$^{-1}$ and 20 km~s$^{-1}$ (see Table~\ref{tab:products}).  We note that the final results are significantly influenced by the choice of key parameters and the ones that primarily affect clump detection are MINDIP, MAXJUMP, and MINPIX (the minimum number of pixels a clump must contain). To minimize the influence of these parameters, and to eliminate obvious noise spikes, we set MINPIX to 16 pixels. We chose FLATSLOPE to be 2 $\times$ rms after several tests showed that no significant differences in the final number of components and the determination of the edges of the components were observed. To mitigate the impact of MINDIP and MAXJUMP, we performed several runs while varying the parameters: [1, 2, 3]-$\sigma_{rms}$ and [4, 6, 8] pixels, respectively. For all runs, the algorithm found a total number of 2 components, which were previously identified as C05-NW and C05-SE. Three components, with 2 located in the C05-NW region, can only be found if the smallest dip in height allowed (MINDIP) is below 1-$\sigma$. This means that the difference between the flux of the two peaks of C05-NW is not significant enough to be considered as two independent components. In conclusion, our analysis favors the scenario that CRISTAL-05 is composed of two components, namely C05-NW and C05-SE.

The Fellwalker algorithm yields a final cube mirroring the size and spectral dimensions of the original, where pixels are assigned positive values as identifiers of each component. In contrast, those not associated with any structure are assigned negative values. We isolate the emission of a given component by masking out the pixels not assigned to that source. For each channel, we apply to the masked pixels a Gaussian distribution with the same rms of the channel. In Figure~\ref{fig:residual-fellwalker}, we display integrated maps and spectra of the whole system after the clump extraction. From left to right, the first panel shows the non-JvM-corrected \cii moment-0 map integrated over the full velocity range covered by the CRISTAL-05 system (same as Figure~\ref{fig:moments}). Green and dashed navy blue contours show the integrated emission of C05-NW and C05-SE extracted by the algorithm, where the major axes of each component are resolved along about 2 and 3 beam sizes, respectively. In the second panel, we present the JvM-corrected spectra of the whole system and each individually identified component. There is significant additional emission not accounted for by C05-NW and C05-SE, especially for velocities greater than 100 km~s$^{-1}$. 

To pinpoint its location and emission distribution, we mask the emission from the two components (C05-NW and C05-SE) out of the original cube and compute the spectrum of the residuals, which is shown in the third panel. In the fourth panel, we see the non-JvM-corrected moment-0 map of the residuals. Most of the compact emission in the central region is removed, and only extended emission is left in the surroundings of the galaxy.

The clump extraction yields a flux of the  C05-NW, C05-SE, and the extended components corresponding to  46\%, 13\%, and 41\% of the total emission of the system, respectively. The components segmentation shows that a substantial fraction of the total \cii line emission comes from the residual emission after subtracting the C05-NW and C05-SE. However, by default, the clump identification is sensitive to what we set as a noise and segmentation threshold, leaving some extended wings behind. The definition of where the central emission ends and the extended third component starts depends on the resolution and sensitivity of the observations.  It could be, for instance, that the two compact components have extended tails on their gas distribution and, because of the lower signal-to-noise, they are not associated with the central regions by the Fellwalker algorithm. Regardless, the emission in the radial profiles of the surface brightness at a distance exceeding a few kpc from the UV effective radius makes clear that there is an extended \cii emission with two bright and compact central components. Considering the points discussed above, we define the fraction of \cii emission in the extended emission as an upper limit, meaning that it contributes no greater than 41\% to the total flux emission of CRISTAL-05.

\begin{figure}
    \centering
    \includegraphics[width=0.4\textwidth]{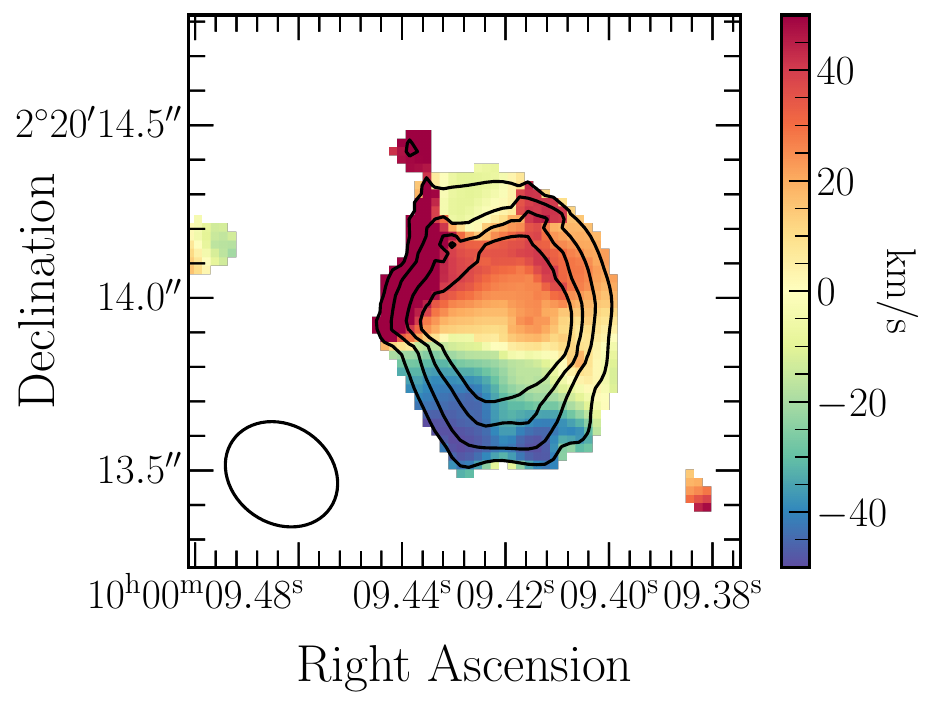}
    \includegraphics[width=0.4\textwidth]{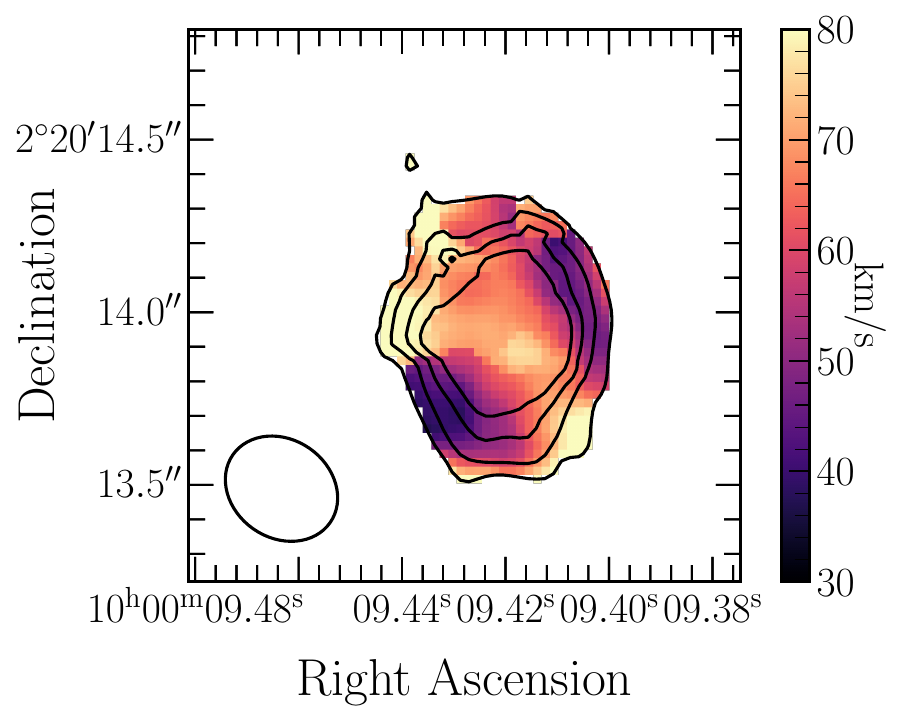}
    \caption{Moment-1 (top) and moment-2 (bottom) maps of C05-NW component, for the frequency with a spectral difference of  -30 km~s$^{-1}$ from the \cii frequency at z = 5.54. The black contours correspond to the 3-, 6-, 9-,
12-$\sigma$ levels of the \cii moment-0 map.}
    \label{fig:CO5-N-kinematic}
\end{figure}

\begin{figure*}
    \centering
    \includegraphics[width=\textwidth]{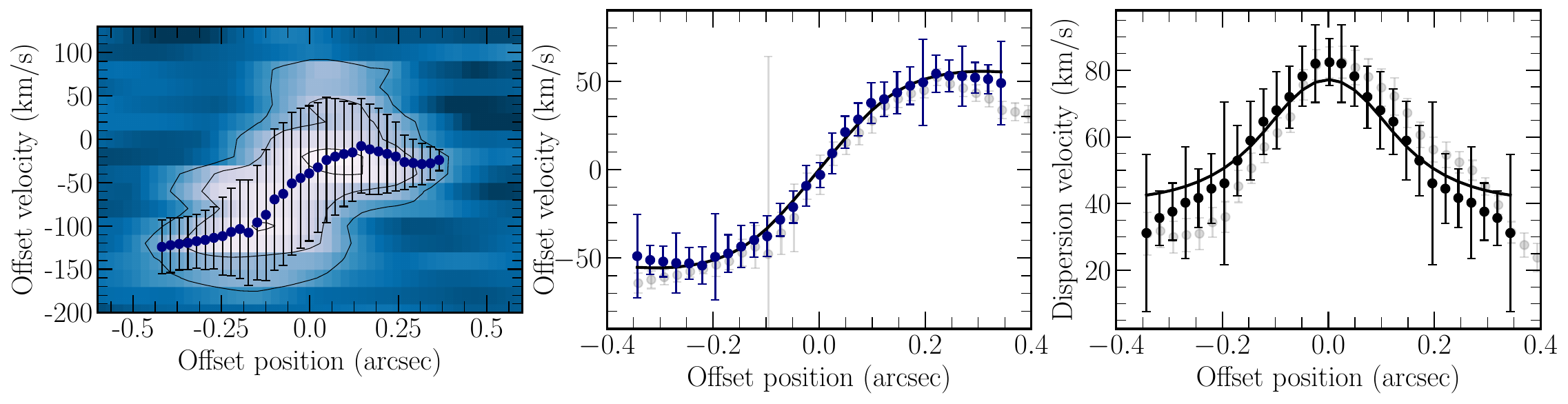}
    \caption{Radial rotation curves of C05-NW as the input for DYSMALpy. (a) Left panel: Position-velocity diagram in an extracted-C05-NW cube, with a slit along the major axis with a width of 0.245\arcsec. The navy circles are the peak velocity of the Gaussian distribution in each position, and the black error bars are the dispersion of the distribution. These curves were symetrized and presented in the central and right panels, respectively. (b) Central panel: Symmetrized rotation radial profile of C05-NW as navy circles. The background gray circles are the not-symmetrized curve, shown as navy circles in the left panel. (c) Right panel: Symmetrized dispersion velocity radial profile of C05-NW as black circles. The background gray circles are the error bars in the left panel before being symmetrized.}
    \label{fig:dysmal-1}
\end{figure*}

\begin{figure}
    \centering
    \includegraphics[width=.45\textwidth]{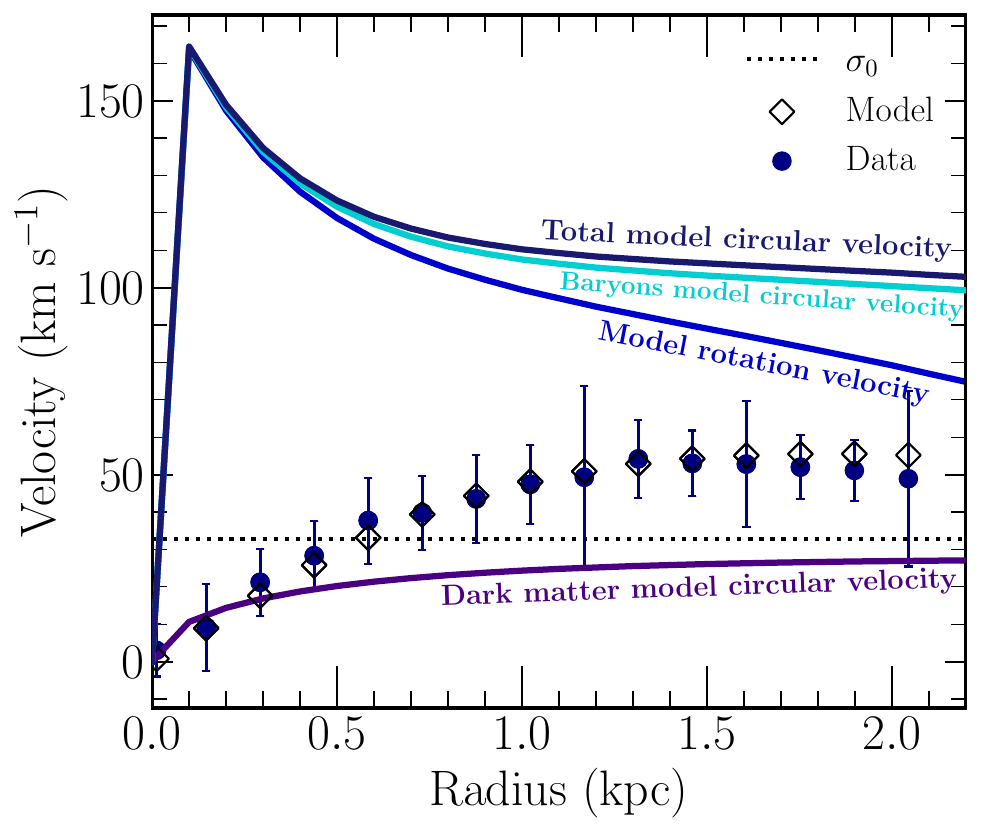}
    \caption{Intrinsic circular velocity and rotation velocity profiles of C05-NW for the best-fit model of DYSMALpy. From top to bottom, we show as solid lines: the intrinsic circular velocity of total mass distribution (baryon + dark matter component, navy color), the intrinsic circular velocity of the baryons (gas + stars, cyan color), rotation velocity curve of the total mass distribution (baryon + dark matter, blue color), the intrinsic circular velocity of the dark matter mass distribution (dark violet color). All the curves are inclination-corrected. As a reference, the intrinsic velocity dispersion is plotted as dotted black lines. The symmetrized rotation curve, the same as Figure~\ref{fig:dysmal-1}, is plotted as navy circles, and the white diamonds show the best-fit model. Both rotation curves are not inclination-corrected.}
    \label{fig:dysmal-2}
\end{figure}

\subsection{Kinematic modeling of C05-NW} \label{sec:kinematical}

In Figure~\ref{fig:CO5-N-kinematic}, we display the moment-1 and moment-2 maps of component C05-NW, extracted from the clump-finding algorithm. C05-NW exhibits signatures of disk rotation, according to criteria commonly used in the literature \citep{2020forster}: (i) a smooth monotonic velocity gradient across the galaxy; (ii) nearly co-location of the kinematical and morphological centers; (iii) co-alignment of the morphological and kinematic major axes; and (iv)  velocity dispersion peaks around the region of the steepest velocity gradient (which is mainly driven by beam smearing), excluding the high dispersion portion in the northeast and southwest of the galaxy. Therefore, we treat C05-NW as a disk in what follows.

We kinematically model the C05-NW emission using the parametric 3D software DYSMALpy \citep{2009cresci,2011davies,2016wuyts,2021price}.  This code employs a multi-component 1D, 2D, or 3D model to compare with the observed cube via Markov Chain Monte Carlo (MCMC) sampling, accounting for observational effects such as beam smearing and instrumental line broadening. We choose the  1-d approach to simultaneously model the radial and dispersion velocity curves along the major axis of the galaxy since it contains most information of the intrinsic rotation curve \citep{2020genzel, 2021price}. 

In the left panel of Figure~\ref{fig:dysmal-1}, we show the position-velocity diagram in the cube masking out the C05-SE and extended emission, with a spectral resolution of 20 km~s$^{-1}$. We created the rotation curves by fitting a Gaussian distribution in each column of the position-velocity diagram. The C05-NW component has features caused by the gravitational instabilities of the interaction: in the negative side of the PV diagram, we see some emission that deviates from the symmetric S-shape around $\sim$ -50 km~s$^{-1}$, while in the positive velocity side, an extra emission extends from 40 to 80 km~ s$^{-1}$. Such regions require either masking or adopting two Gaussian fittings in a giving column. The peaks from the Gaussian fitting, which correspond to the velocity of the rotating curve, are plotted as navy circles, and the error bars correspond to the dispersion of each column. 

In the central and right panels of Figure~\ref{fig:dysmal-1}, we show the line-of-sight and dispersion velocity curves (in navy and black points, respectively), used as an input to DYSMALpy. Both curves were symmetrized by averaging the velocity values as a function of the distance from the galaxy's center. This was motivated by the observed asymmetries beyond 0.2\arcsec in both curves (see PV diagram Figure~\ref{fig:dysmal-1}, left panel). The zero point of both offset position and velocity were shifted to correspond to the symmetry point of the curve and better represent the kinematical center of the component. The velocity peaks at 0.2\arcsec and C05-NW is very compact, with an effective radius of $\sim$ 700 pc ($\sim$ 0.12 \arcsec), meaning that the curves extend up to $2.5\times{\rm R_{eff}}$. %The full extent of the 1D kinematic profiles is covered by $\sim3$ resolution elements.

To perform the kinematical fitting, the mass model consists of a thick disk, a bulge, and a dark matter halo. The parameter selection is summarized as follows:

a) Dark matter mass distribution: It follows a Navarro-Frenk-White profile \citep{1996navarro}, with a fixed concentration as c = 3, which is expected at this redshift \citep{2014dutton}.  We let the dark matter fraction (fdm) within the effective radius be a free parameter, with an initial guess of 0.5 and a flat prior. The virial mass is tied to the fdm, centered at $\rm log \, M\rm_{DM,vir}/M_{\odot} = 12.1$, sigma of 0.5 and bounded at $ \rm log \,  M\rm_{DM,vir}/M_{\odot} =$ [9, 13]. We do not assume an adiabatic contraction of the halo.

b) Baryon mass distribution (disk+bulge component):

We set the total baryonic mass as a free parameter with an initial guess of log M$\rm_{bar}$/M$_{\odot}$ =  10.5, by the sum of the stellar mass log M$_{*}$/M$_{\odot}$ = 10.27 obtained by \cite{2015capak}, and the assume a gas-mass ratio of log M$\rm_{gas}$/M$_{*}$ = 0.75 \citep{2020dessauges}. We choose a truncated Gaussian prior, centered at log M$\rm_{bar}$/M$_{\odot}$ =  10.5, with a standard deviation of 0.5 and varying in a wide range M$\rm_{bar}$/M$_{*}$ $\in$ [9.5,11]. This choice of bounds is based on the fact that the stellar mass estimation assumes CRISTAL-05 is a single source since it is unresolved in the HST imaging \citep{2015capak}.

Our choice of a disk+bulge mass modeling is motivated by the failure of initial tests in fitting just a disk component, which can not fit the central peak dispersion velocity of the galaxy.  We assume a fixed bulge-to-total ratio of 30\%. The disk component follows a Sersic profile with a fixed Sersic index n = 1 and the effective radius is left free with a truncated Gaussian prior varying in [0.5,3] kpc, centered at 1.32 kpc and standard deviation of 1 kpc.  For the bulge component, we fixed the Sersic index and effective radius as n = 4 and r$\rm_{eff}$ = 0.15 kpc. We fix the intrinsic thickness of this component to q$_{0} =1$. The thick disk requires an asymmetric drift correction following \citet{2010burkert}. Finally, we assume that the dispersion velocity is isotropic and constant along the galaxy, so we add an initial guess of 40 \kms with a flat prior varying [5,100] \kms. 

(c) Geometry: We fixed the kinematic center and position angle based on the moment-0 \cii emission map. The inclination was set to 60°, according to the axial ratio of a  spatial Gaussian fitting in the moment-0 map.

In summary, the free parameters are the dark matter fraction, disk effective radius, and dispersion velocity. We ran the algorithm using 1000 walkers, a burning phase of 50 steps, a running phase of 200 steps, and a final average acceptance fraction of 0.3. This led to the following maximum a posteriori estimates of our observations (by jointly analyzing the posteriors of all free parameters, Appendix~\ref{appendix-dysmal}): log M$_{bar} = 9.83^{+0.11}_{-0.17}$ M$_{\odot}$, r$_{disk}^{eff} = 1.82^{+0.92}_{-0.52}$ kpc, $M_{DM,vir} = 8.85_{-0.33}^{+1.95}$, dark matter fraction within the effective radius fdm = 0.06$^{+0.27}_{-0.06}$ and $\sigma_{0} = $ 35$_{-7}^{+6}$ \kms, indicating the range within which we have 68\% confidence that the true value lies.  

In Figure~\ref{fig:dysmal-2}, we display the intrinsic circular velocity curves of the total, baryon, and dark matter components. The intrinsic circular velocity peaks at 100 pc, with a velocity of 155 \kms, and it remains flat at r $>$ 0.6 kpc for a value of 120 \kms. The dark matter contribution remains small at all radii probed by the \cii observations (r $<$ 2 kpc).  This is a good fit for the velocity curve ($\chi^{2}$ = 0.63) but moderate for the dispersion velocity curve ($\chi^{2}$ = 2.6).  The kinematical modeling retrieves a rotating gaseous disk in a compact source, mainly baryon-dominated. At distances greater than the beam size, the rotation-to-dispersion is V/$\sigma_{0} \sim$ 3, which implies dominant rotational support. However, the main axis of the galaxy is resolved over $\sim$ 3 beam sizes, therefore we can not completely rule out that C05-NW is actually a compact merger whose nuclei are closer than 2 kpc.

\section{Discussion}

\subsection{On the nature of CRISTAL-05}

Our observations of CRISTAL-05 can provide insights into the physical mechanisms that have affected the mass assembly of galaxies. Whilst having a compact rest-UV and -NIR emission, CRISTAL-05 has a gaseous disk component with features of disturbance. The multicomponent spectra and morphology indicate either that C05-NW and C05-SE are internal clumps of a single galaxy or two independent galaxies in a close approach. The nature of these structures will be discussed in this section.

The first scenario states that the complex morphology and kinematics of CRISTAL-05 reassemble a single galaxy hosting massive star-forming clumps. This clumpiness can be formed in situ from gravitational fragmentation in gas-rich disks with relatively high turbulent speeds \citep{2008bournaud,2008genzel, 2008elmegreen, 2010ceverino, 2014dekel}. Galaxies at such redshift are gas-rich and constantly fed by cold gas coming from the intergalactic medium, which is the main driver of the turbulence \citep{2009dekel, 2010ceverino,2014dekel, 2016rodriguez-gomez}. An alternative origin is an ex-situ formation, where they are remnants of merging satellite galaxies \citep{2017ribeiro, 2024nakazato}.   

Indeed, bright substructures can imprint emission bumps in the \cii spectrum of a galaxy \citep{2019kohandel,2020kohandel}, explaining the heavy-tail emission in Figure~\ref{fig:spectrum-cii}.  In addition, the spatial proximity of the peak of the rest-UV and rest-NIR emissions with the \cii centroid shows that, in this scenario, they trace the nucleus of  CRISTAL-05 galaxy, with the star-forming regions being slightly more extended than the older stellar component (Figure~\ref{fig:radialprofile}, top panel).

To date, direct detections of giant clumps within normal galaxies have primarily been observed in massive rotating disks during the cosmic noon period (1 $<$ z $<$ 3). These clumps exhibit sizes reaching up to 1 kpc and typically possess stellar masses no greater than M$_* \sim 10^{9}$ \citep{2008genzel}. Beyond redshift z $>$ 4, the identification of such clumps is scarce, primarily found within massive starburst systems \citep{2022spilker} and lensed star-forming galaxies \citep{2022mevtric,2022vanzella, 2024messa,2024fujimoto}. 
%These instances are often attributed to gravitational interactions with nearby companions and the fragmentation of disks driven by cold streams.
Our understanding of the clumps in typical galaxies also relies on simulated observations of star-forming regions within galaxies at redshifts 5 $<$ z $<$ 7, such as those provided by the SERRA and FIRE hydrodynamical simulations \citep{2022pallotini, 2023wetzel}. Some examples of galaxies in the SERRA simulation show typical clump sizes of $\lesssim$ 0.5 kpc \citep{2020kohandel, 2021zanella, 2022pallotini}. 
The spatial Gaussian fitting of the \cii in  C05-NW and C05-SE yields deconvolved sizes of  $0.4 \times 0.2$\arcsec ($2.4 \times 1.5$ kpc) and $0.22 \times 0.16$\arcsec ($1.3 \times 0.9$ kpc), respectively. 
Although these values may represent lower limits for the true sizes, they still exceed what would typically be expected for clumps at this redshift and in the cosmic noon.
Additionally, if we consider that the UV contribution of C05-SE is negligible, the effective radius of C05-NW (0.3 kpc) is at least an order of magnitude large compared to stellar clumps recently identified in lensed galaxies \citep{2022mevtric,2022vanzella, 2024messa,2024fujimoto}.  We can not perform the same exercise for C05-SE, so deeper UV observations are required. 
Lastly, we note that the clumpy scenario struggles to account for the distinct velocity distributions observed in the position-velocity (pv) diagram. Notably, there is no observable emission stream connecting the components, and a substantial difference of 150 km s$^{-1}$ and 100 \kms exists between the negative components of C05-NW and C05-SE, and the central portion of C05-NW and C05-SE, respectively. %horri

According to mock observations the \cii, rest-UV, and rest-NIR emission of the typical galaxy \textit{Althaea} at z $\sim$ 6 simulated by the SERRA suite, it is required a resolution of about $\theta \sim 0.05$\arcsec to detect substructures within the galaxy \citep{2021zanella}. Therefore, our limited resolution to detect internal clumps and the difficulty of this scenario to explain the disturbed features in the shape and dynamics of the gas point to CRISTAL-05 being a merging system. We hence presume that  C05-NW and C05-SE are two galaxies, originally formed apart, in a close approach ($\sim$ 2 kpc) and thus undergoing a merging event. The spatial projected proximity ($\sim$ 2 kpc) and velocity difference of $\sim$ 100 \kms show that the system is gravitationally bound. The gravitational interactions distort and cause the gas stripping to the outskirts of the galaxies, as we can see in the disturbed gas imprinted in the \cii moment-0 map and pv diagram (Figure~\ref{fig:arm-pv-spec}). The higher brightness and size of C05-NW points it as the dominant, most massive galaxy of the system. The compactness of the UV and NIR emission is most likely attributed to the sensitivity of the current observations that fail to detect faint UV and NIR emission of satellite galaxies \citep{2021zanella}. Another possibility is that the C05-SE galaxy is heavily dust-obscured, as the dust emission peak coincide with this region. 

%We are witnessing the mass assembly of a typical galaxy in the first billion years of the universe. Such events are expected to be frequent 

%Galaxy mergers are also thought to play an important role in the stellar mass growth over cosmic time

%(i) higher merger rates (e.g. Conselice 2006; Wetzel
%et al. 2009; Lotz et al. 2011) and (ii) generally higher gas frac-
%tions (e.g. Santini et al. 2014; Dessauges-Zavadsky et al. 2017;
%Schinnerer et al. 2016; Scoville et al. 2017; Tacconi et al. 2018),
%and might also explain previous similar observations of extended
%[CII] emission around multi-component systems at z ∼ 5 (s

Finally, we follow a simple approach to calculate the dynamical mass of C05-NW and C05-SE by M$_{\rm dyn}$ = $\rm\gamma$Rv$^{2}$/G in units of M$_{\odot}$ \citep{2015spilker,2022spilker}, where G is the gravitational constant ${\rm G}=4.32\times10^{-6}$ \kms kpc M$_{\odot}$, R is the radius (in units of kpc), v the velocity of the system in \kms and $\rm\gamma$ is the pre-factor which accounts for the geometry of the system. The radius R is assumed to be the circularized radius from a beam-corrected Gaussian fitting on the moment-0 map of each component. For C05-NW,  we assume the velocity v = 120 \kms at R = 1.5 kpc, where the rotation curve, retrieved by the kinematical modeling, is flat (Figure~\ref{fig:dysmal-2}). For C05-SE, the system is dispersion-dominated (see pv diagram), so we assume $\gamma$ = 5 \citep{2022spilker} and v $\sim \sigma$ of the system. It results in a dynamical mass of M$_{\rm dyn}^{\rm NW} = 5 \times 10^{9}$ M$_{\odot}$, M$_{\rm dyn}^{\rm SE} = 9.1 \times 10^{8}$ M$_{\odot}$, which are consistent with the value estimated for C05-NW in section \ref{sec:kinematical} and the dynamical mass of typical galaxies at z = 5.5 \citep{2022spilker}.
It leads to a mass fraction of M$_{\rm dyn}^{\rm SE}$/M$_{\rm dyn}^{\rm NW}$ = 0.2, consistent with a minor merger. 
%Assuming a fiducial gas fraction of 50\% , we can estimate that the stellar mass of \NW is $\sim2.5 \times 10^{9}$ M$_{\odot}$, which is in the middle of the range of stellar masses measured at similar redshifts \citep{2015Grazian,2016Song,2023Navarro_Carrera}.

\subsection{\cii extended emission}

Several studies in the literature refer to \cii haloes in early galaxies as the existence of carbon-enriched gas surrounding the galactic disk where most of the star-forming activity takes place. The typical sizes of 6-12 kpc \citep{2020fujimoto,2021herrera-camus,2022akins} of these gaseous components exceed the effective radius of UV emission \citep[1-3 kpc,][]{2020fujimoto}, reaching the circumgalactic medium, loosely defined as the gas beyond 1 - 2 $\times$ the effective radius of the galactic disk out to the virial radius. It suggests that the circumgalactic medium is also composed of a cold gas phase, in addition to the usual hot and warm gas \citep[][and references in]{2017tumlinson}, as a result of galactic-scale processes such as nuclear, and stellar outflows and intergalactic gas accretion. 

It is well known that the \cii emission is associated with the star formation activity of local and distant galaxies \citep{2014delooze,2015herreracamus,2018carniani} since this is the major cooling line of stellar heating, in different gas phases \citep{2010stacey}. The lack of heating sources to generate this emission at far distances requires some physical mechanism to transport the carbon produced by the stellar nucleosynthesis to the outskirts and to constantly excite the singly ionized phase.

Currently, there are different approaches to identifying a \cii halo: (i) the first one was adopted by \citet{2020fujimoto}, where a galaxy contains a \cii halo when R$^{eff}_{[CII]} >$ R$^{eff}_{UV}$ and it has a \cii intensity in the outskirts of the galaxy greater than 4-$\sigma$ for \cii, but below 3-$\sigma$ for the UV continuum. This extended area is delimited by an aperture radius of 10 kpc, masking out the emission within a beam area centered at the source peak emission. The outer radius is motivated by the spatial scale found via the visibility-based stacking of 18 star-forming galaxies at
z = 5–7 by \citet{2019fujimoto}. (ii) The second one, adopted by kpc-resolved studies \citep{2021herrera-camus,2022akins} and also quantified by \citet{2020fujimoto}, claims a \cii halo when R$^{eff}_{[CII]} >$ R$^{eff}_{UV}$ and there is a significant intensity gap between the normalized surface brightness radial profiles of \cii and UV emission, beyond the noise level. 
(iii) The third approach, performed by \citet{2020ginolfia} and \citet{2024dicesare}, applies to major mergers (velocity separation of $\rm\Delta v \le 500 \, km s^{-1}$) with a clear projected separation ($\rm r_{p} \ge 4$ kpc). The identification of the halo occurs when the emission is detected within the region enclosed by the 2-$\sigma$ contour of the \cii moment-0 map after subtracting the emissions from each interacting galaxy. The emission of each component is accounted for within the region defined by the FWHM of the Gaussian fit to the \cii spatial emission.  These studies not only detect the emission but also attribute at least 50\% of the total emission to these \cii envelopes.

According to definitions (i) and (ii), \cite{2020fujimoto} defined CRISTAL-05 as having a \cii halo. Since the first approach depends on the relative sizes of the galaxy and the beam, and the third one can be not applied to CRISTAL-05 (blended emission with a projected separation of $\rm r_{p} \sim 2$ kpc), we stick to the second one which evaluates the relative morphologies and extension of the different emissions.  The higher resolution and depth of our observations confirm the extended nature of \cii emission compared to both the rest-UV, at least up to 10 kpc. The detection of an extended \cii emission so far assumes that the rest-UV emission from HST observations is a good tracer of the galactic disk because this is where most of the star formation takes place. However, rest-frame UV observations can be affected by dust obscuration and thus not necessarily trace the galaxy star formation distribution \citep{1999meurer, 2005burgarella}.
Moreover,  the star-forming regions tend to be more compact and it does not necessarily trace the mass distribution of a galaxy \citep[][]{2022pallotini, 2022ferreira}. 
Thus, the rest-UV emission is not enough to specify the galactic disk extent, and therefore, information on the stellar spatial distribution is also required. Hence, we also added the available JWST/F770W-band imaging of CRISTAL-05, which at z = 5.54 traces the emission close to the peak of the continuum of older stellar populations at z=5.54 (rest-frame 1.2 um). Similarly to the UV-continuum, the stellar emission is compact, reinforcing the extended nature of \cii emission in CRISTAL-05. % (R$^{eff}_{F770W}$ $<$ R$^{eff}_{\cii}$)

We thus describe the main physical mechanisms proposed in the literature to explain the origin of the extended \cii  emission \citep{2019fujimoto, 2020ginolfia, 2022akins}:

\begin{enumerate}
    \item Satellite galaxies - a non-negligible number of low-mass galaxies reside spatially around the massive galaxy and the \cii emission arising from stellar heating in these smaller systems is smeared by the current resolution of observations, therefore mimicking an extended emission. 
    \item Cold streams - intergalactic gas streams flow toward the potential well of the central massive galaxy. \cii emission is the product of gravitational energy and/or shock heating. 
    \item Outflows - high-speed interstellar gas flows outwards the potential well of the massive galaxies as a product of AGN and/or stellar feedback. 
    \item Bridges and tidal tails associated with mergers - the cooling of \cii release emission via kinematic energy dissipation in a turbulent cascade from large to small scales. 
    %\item Extended disk - 
\end{enumerate}

\begin{figure*}
    \centering
    \includegraphics[width=0.37
\textwidth]{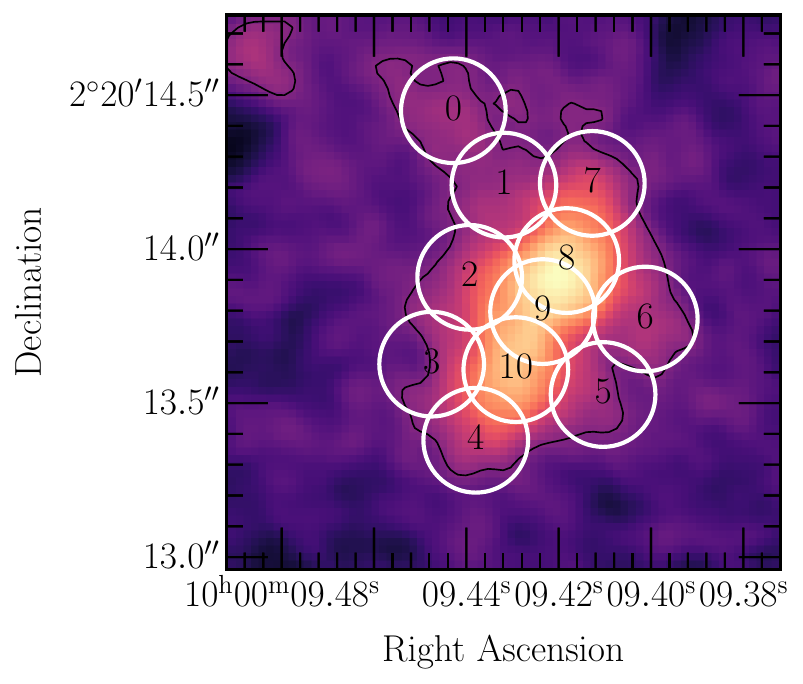}
    \includegraphics[width=.6
\textwidth]{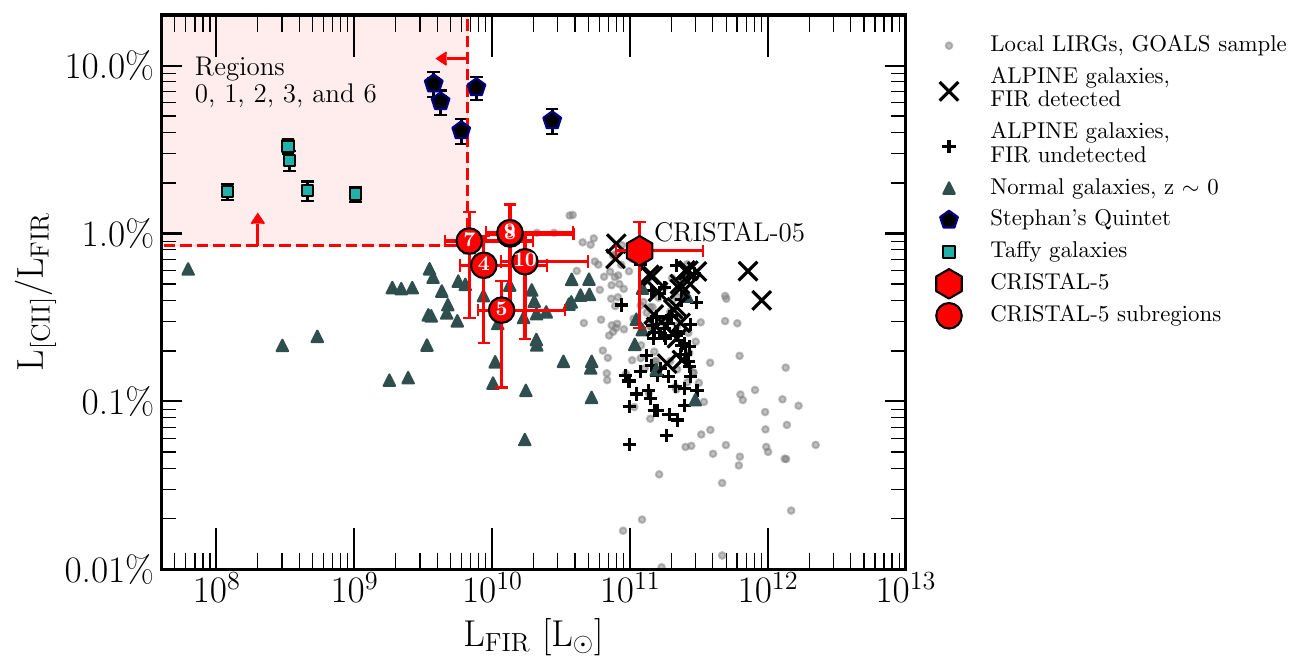}
    \caption{\textbf{(a) Left panel:} \cii moment-0 map, same as Figure~\ref{fig:moments}, with 0.17\arcsec-apertures  across the galaxy (white circles). The emission extracted is analyzed in the right panel and Figure~\ref{fig:discussion-resolved-cii2}. \textbf{(b) Right panel:} \cii-to-FIR luminosity ratio, as a function of the FIR luminosity. The global measurement (aperture of 1.5\arcsec) is shown as a red hexagon, and the subregions 8, 9, and 10 are shown as red circles. The uncertainties reflect the possible dust temperature range between 40 and 60 K. The red area locates the upper limits of the mean of the ratios, where the dust emission is undetected. As a comparison, we also reference to z $\sim$ 0 LIRGS of the GOALS sample \citep[gray circles;][]{2013diazsantos}, z $\sim$ 0  normal galaxies \citep[darkgray triangles;][]{2001malhotra}, a purely shock-gas region in the Stephan's Quintet compact group \citep[navy pentagons;][]{2013appleton}, shock-gas in the bridge of the merging pair UGC12914/12915 \citep[cyan squares;][]{2018peterson}, and 4 $<$ z $<$ 6 typical galaxies from the ALPINE galaxies \citep[IR detected sources as x-shape marker and IR upper limits as plus-shape markers;][]{2020bethermin}.  }
    \label{fig:discussion-resolved-cii1}
\end{figure*}

Our high-resolution observations provide evidence of the merger nature of CRISTAL-05, supporting scenario (4) as the main driver for the presence of \cii emission beyond $\sim$ R$_{UV}^{eff}$. Simulations point out that satellites in minor mergers can disrupt the gas in low-mass galaxies \citep{2020kohandel, 2022akins}. This result impacts the detection and analysis of the current properties of galaxies with \cii halos. In the ALPINE survey, \citet{2020fujimoto} considered only galaxies classified as non-mergers to avoid the obvious \cii emission caused by shock heating in tidal stripping gas. Following this criterion, they concluded that 30\% of the single ALPINE sources are classified as containing a \cii halo ($>4$-$\sigma_{[CII]}$ and $>3$-$\sigma_{UV}$ in the outskirts of the galaxies). Comparing both samples, galaxies with haloes are more massive and starbursting, and have most likely past/ongoing stellar-driven outflow episodes to form this kind of haloes \citep{2020ginolfib}. 
A similar analysis by \citep{2019fujimoto} could not reproduce the \cii excess in surface brightness radial profiles of the stacked observations of the ALPINE survey in state-of-art cosmological simulations \citep{2017pallottini, 2017yajima}. 
Our results suggest that at least a fraction of the galaxies with a \cii halo that are located in the bright end of the SFR-M$_{*}$ relation could be multi-component systems misclassified as individual galaxies. In such cases, the spectral energy distribution (SED) fitting on the non-resolved sources took into account the summed photometry of the system that can mimic a more massive system. For these sources, the tension between observational and cosmological simulation breaks since we are analyzing objects with different dynamical natures.

This outcome highlights the need for high-resolution observations to ultimately distinguish isolated galaxies from mergers, requiring at least three beams along the major axis and an S/N $>$ 10 \citep{2022rizzo}. A close encounter, such as CRISTAL-05, imprints a low-velocity gradient in moment-1 and position-velocity diagrams, requiring a higher spatial and spectral resolution to efficiently separate two or more components.

Other examples in the literature also support merger activity as the cause of \cii extended emission. The typical star-forming galaxy HZ7 at z = 5.25 contains an extended \cii emission \citep{2023lambert} with no evidence of a broad secondary component in the \cii spectrum across the galaxy, that could point to outflows. Instead, the disturbed integrated gas and dust emission and velocity field points to a late-state merger. Similarly, the strongly lensed sub-L* galaxy A1689-zD1 at z = 7.13 also has a \cii halo, with signs of being a multicomponent system \citep{2022akins}. 

As previously mentioned in the third definition,  \citet{2024dicesare} analyzed carbon-rich envelopes of 6 major mergers of the ALPINE survey and simulated galaxies at z = 4 - 5. The study concludes that the gas between and beyond the individual galaxies constitutes more than 50\% of the total \cii emission.  These extend across distances exceeding 20 kpc (average 27 kpc), while isolated systems with extended emission from ALPINE survey go up to 10 kpc. They also found that the fraction of \cii in the envelope correlates with the projected separation of the systems and the mass ratios, meaning that more mergers in advanced stages and with masses ratios close to unity host stronger gravitational interactions that can more intensively disrupt the gas of the galaxies. The low fraction ($<$ 40\%) and extension of CRISTAL-05 point that the galaxy is a faint analog of such systems. 

% \subsubsection{Exploring the [CII] to IR ratios}

\subsection{Shock heating exciting the extended \cii emission}

The \cii arises mostly from the gas in photodissociation regions via photoelectric heating from polycyclic aromatic hydrocarbons (PAHs) and small dust grains \citep{1972watson, 1978draine, 1985tielens,1994bakes, 2010stacey,2012croxall}.  When considering thermal equilibrium, a clear correlation between \cii emission and star formation activity emerges, applicable to both local and high-redshift galaxies \citep{2014delooze,2015herreracamus, 2018carniani, 2020schaerer}. Nonetheless,  other mechanisms can also contribute to \cii emission. For instance, an excess in the \cii/FIR ($> 1$\%) and \cii-to-PAH ratios have been indicative of low-velocity shocks \cite{2013appleton, 2013lesaffre}, observed in galaxy mergers \citep{2018peterson}, compact groups \citep{2014alatalo, 2017appleton, 2023fadda} and infalling cluster galaxies \citep{2022minchin}. Since the dust continuum emission predominantly arises from star formation activity, an upward deviation from these ratios may indicate that PDR heating is not enough to explain an enhanced \cii emission.

We explore this scenario by evaluating how the \cii emission varies relative to the far-infrared emission (FIR) across the galaxy. We define apertures with the same width of the beam size (${\rm d}\sim0.33$\arcsec $\sim$ 2 kpc), shown in the left panel of Figure~\ref{fig:discussion-resolved-cii1}. As we can see in the \cii moment-0 map, the apertures isolate the different features of the galaxy: regions 0 to 7 comprise the area around the central sources, with region 0  tracking the UV-dark emission in the northeast arm. Regions 8 to 10, on the other hand, sample the area in and between the two galaxies. Emissions from regions 8 and 10 arise mostly from C05-NW and C05-SE, respectively, although contamination from disturbed gas in the same line of sight cannot be excluded.  

For the FIR luminosity estimation, we convert the dust continuum fluxes at the rest-frame 158 $\mu$m using the scaling factor derived by \citet{2020bethermin}. It describes the average SED for typical galaxies at z $>$ 4, assuming a dust temperature of T$_{d} = 45$ K and an emissivity index of $\rm\beta_{dust} = 1.5$. For the FIR, which accounts for the emission in a wavelength coverage between 42 $\mu$m and 122 $\mu$m, the conversion factor is $\nu$L$_{\nu}$(158 $\mu$m)/L$\rm_{FIR}$ = 0.185 \citep{2024solimano}. The uncertainties reflect the possible dust temperature range between 40 and 60 K, also obtained by the available conversion factor in \citet{2020bethermin}. The global emission (circular aperture of r = 1.5\arcsec) is displayed as a red hexagon in Figure~\ref{fig:discussion-resolved-cii1}, and the FIR emission is detected in regions 4, 5, 7, 8, 9, and 10 (red circles). For the non-detection regions, the red area corresponds to the mean of upper limits ratios, also assuming T$_{d} = 45$ K.

We compare our results with other galaxies in the literature:  z $\sim$ 0 LIRGS of the GOALS sample \citep[gray circles;][]{2013diazsantos}, z $\sim$ 0  normal galaxies \citep[darkgray triangles;][]{2001malhotra}, and 4 $<$ z $<$ 6 typical galaxies from the ALPINE sample \citep[we also apply the conversion factor of $\nu$L$_{\nu}$(158 $\mu$m)/L$\rm_{FIR}$ = 0.185,][]{2020bethermin}.  These galaxies have typical \cii/FIR ratios of normal star-forming galaxies \citep[0.1-1\%,][]{2015cormier,2017smith,2018herreracamus} and starburst galaxies \citep[0.1-0.01\%,][]{2001malhotra,2013diazsantos,2018herreracamus}.

We also add to  Figure~ \ref{fig:discussion-resolved-cii1} (right panel)  two cases of atypical \cii/FIR ratios. \citet{2013appleton} found an excess in the and \cii-to-FIR ratio in a pure-shock region of Stefan's Quintet (SQ), arguing against purely photoelectric-heated gas in photodissociation regions. 
The second case is identified in the bridge between the local merging Taffy galaxies \citep[UGC 12914/12915][]{2018peterson}. For both systems, the \cii luminosity reaches 2-10\% of the total FIR emission, which contrasts with the fraction in normal galaxies and local LIRGs nuclei \citep[darkgreen triangles and grey circles,][]{2001malhotra, 2013diazsantos}.

The global measurement of CRISTAL-05 places it around the upper envelope of the location of local LIRGs. Regions 4, 7, 8, 9, and 10, which comprise the regions in and between C05-NW and C05-SE, have a luminosity ratio of $\lesssim$ 1\%, similar to normal galaxies (gray triangles). In contrast, we note that the upper and lower limits on the L$\rm_{FIR}$ and L$_{\cii}$/L$_{FIR}$ values, corresponding to the regions around the two main galaxy components of CRISTAL-05 (0, 1, 2, 3, and 6), are consistent with those found in the overlap/shocked regions of the SQ and Taffy systems.

\begin{figure}
    \centering
     \includegraphics[width=0.5\textwidth]{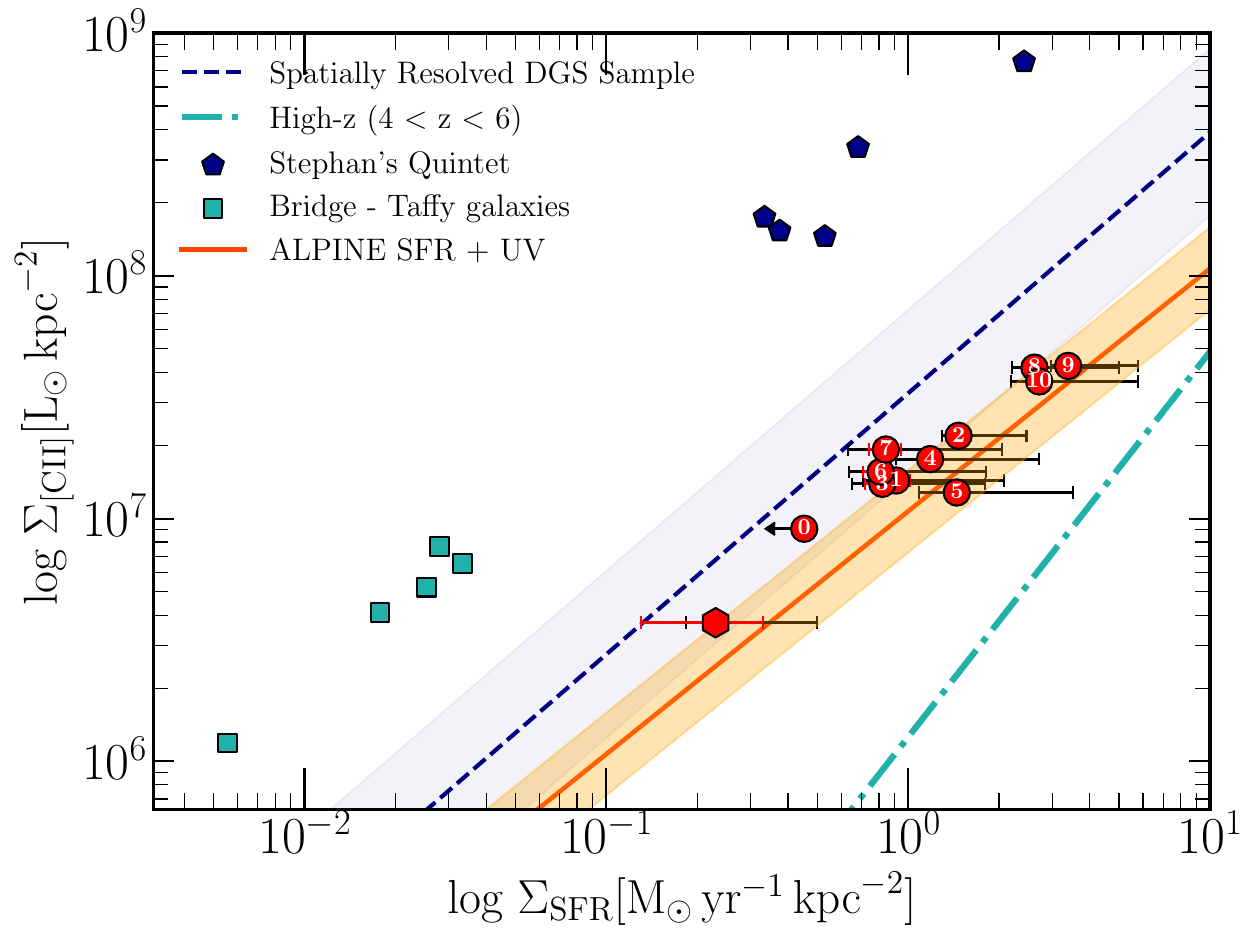}
    \caption{\cii surface density of CRISTAL-05 and its subregions as a function of the surface density star formation rate. The global measurement (aperture of 1.5\arcsec) is represented as a red hexagon, and the subregions as red circles, labeled as shown in the left panel Figure~\ref{fig:discussion-resolved-cii1}. We also display relations of local \citep[navy dashed line,][]{2014delooze} and high-redshift galaxies  (z $>$ 4), as cyan dotted-dashed \citep{2018carniani} and navy dashed \citep{2020schaerer} lines. For the latter, we convert the L$\rm_{[CII]}$ - SFR correlation to the surface density, assuming an average area of 8 kpc$^{-2}$.  }
    \label{fig:discussion-resolved-cii2}
\end{figure}

We thus investigate if we can identify in the system a positive deviation in the $\Sigma_{\rm L_{[CII]}}-\Sigma_{SFR}$ relation which may point to an enhancement of the \cii emission caused by an additional mechanism rather than photoelectric heating.  Although high-z galaxies follow the local L$_{\cii}$-SFR relation, they show a \cii deficiency in the $\Sigma_{L_{\cii}}-\Sigma_{SFR}$ relation compared to local ones \citep{2018carniani}, due to their starburst nature, low metallicity or hardness of the UV radiation field \citep{2019ferrara}. This discrepancy is shown in the dotted-dashed cyan lines in Figure~\ref{fig:discussion-resolved-cii2} \citep[galaxies at 4.5 $<$ z $<$ 7,][]{2018carniani}, compared to the local relation (solid navy color line).  We also add the relation for the ALPINE survey obtained by \citet{2020schaerer}. To make this comparison, we converted the L$_{\cii}-$SFR relation \citep[Figure 2 of][]{2020schaerer} to the surface density one assuming an area of 8 kpc$^{2}$, comparable to the effective radius of the \cii and FIR emission. 
%A similar deficiency is seen for local HII galaxies and high-z galaxies (z $>$ 1), as orange dashed and black dotted lines \citep{2014delooze}, respectively. To make this comparison, we converted the L$_{\cii}-SFR$ relation of these HII galaxies and high-z galaxies to the surface density one assuming an area of 7 kpc$^{2}$, comparable to the effective radius of the \cii and UV emission. 
The shock regions in the Taffy galaxies and SQ, derived by \citet{2018peterson} and \citet{2013appleton} respectively, clearly lie above the local relation. We see that the \cii emission in both the global and most subregions (0, 1, 2, 3, 6, 7) is enhanced compared to galaxies at z $>$ 5  both derived by \citet{2018carniani} and \citet{2020schaerer}, e.g. we do not see a \cii deficiency.

The boosted \cii emission in the regions around the galaxy when compared to galaxies at the same epoch, and the excess in the \cii/FIR ratio are tentatively consistent with a scenario where the  \cii emission is a consequence of heating produced in purely-shock regions of mergers, like what is seen in the Taffy and SQ systems. In fact, \citet{2020Zhang} showed that galaxy interactions can increase some emission line ratios which are consistent with the shock produced by mergers.  Additionally, studies of local galaxies and simulations have also revealed that mergers of galaxies can produce extended gas components without extended stellar distributions as a consequence of gas accretion or collisional processes, such as ram pressure stripping and shocks  \citep{2022Sparre}. In the particular case of CRISTAL-05, the heating produced in the merger may not be as powerful as it happens in the local analogous, possibly due to the low mass of the companion satellite galaxy.  

However, we can not ultimately rule out the occurrence of other physical mechanisms that contribute to the enhanced \cii emission. 
%Infalling galaxies in clusters  have shown enhanced \cii emission caused by ram pressure stripping \citep{2022minchin, 2022Sparre, 2023fadda}. 
%Although \source does not seem to be part of an overdensity (Appendix~\ref{appendix-overdensity}), and we do not expect an intracluster in place at such a redshift, we can not completely exclude this scenario. 
For instance, past/ongoing stellar feedback can contribute to the construction of the extended \cii emission and metal enrichment of the circumgalactic medium of z $\sim$ 5 galaxies.  
Recent semi-analytical models reproduce the extended \cii radial profiles by including starburst-driven, catastrophically cooling outflows \citep{2020pizzati, 2023pizzati}. In this scenario, supernova explosions expand in velocity ranges of 200-500 \kms outwards the galaxy, and the gas undergoes a catastrophic cooling in the first kpc. It is slowly heated up by the cosmic UV background, and its expansion is stopped around 10 - 15 kpc by the gravitational field of the host halo gravitational field. Stacking data in the ALPINE sample have hints of a broad Gaussian feature in the \cii spectrum, most likely caused by starburst-driven outflows \citep{2020ginolfib}. For CRISTAL-05, the merger can enhance the star formation activity, indicating that more frequent supernova events can still be in place. Observations of lines tracing the ionized gas are necessary to identify the contribution of stellar feedback to the circumgalactic medium of interacting galaxies with extended \cii emission.

We conclude that deeper studies are needed to detect the faint dust emission in the outskirts of CRISTAL-05 and to constrain dust temperature. Additionally, more detailed spectroscopy of optical lines is necessary to ultimately define the role of shocks in producing the extended \cii emission of merging galaxies in the early universe, as well as to investigate the contributions of metallicity and other gas phases to the production of \cii.

\section{Conclusions}

We present sensitive, kpc-scale resolved observations of the \cii line emission in the massive, main-sequence star-forming galaxy CRISTAL-05 at z $\sim$ 5.54. Based on these \cii observations from the ALMA CRISTAL large program (plus previous archival and pilot observations), HST/WFC3 rest-frame UV and JWST MIRI rest-frame optical imaging, we conduct a detailed morphological and kinematic analysis of this system. Previous studies showed this is a typical star-forming galaxy, with \cii emission extending beyond the galactic disk, which classifies it as a \cii halo \citep{2020fujimoto}. We summarize our main findings as follows:

\begin{enumerate}
        \item \textbf{Morpho-kinematic classification based on high-resolution data:} The \cii emission shape is elongated (southeast to the northwest direction), as a result of a double peak distribution with project separation of 2 kpc, and a velocity difference of $\sim$ 100 \kms. The full-system velocity field is irregular, and the velocity dispersion is roughly constant around $\sim$ 80 \kms. All these features indicate that CRISTAL-05 is actually a multicomponent system that previous low-resolution observations could not clearly identify. The proximity and low-velocity range of [-50,+50] \kms requires high-resolution observations to properly perform a morpho-kinematical classification.
        \item \textbf{A complex kinematical structure:} A more detailed examination shows that the CRISTAL-05 system is comprised of a brighter NW component with disk-like kinematics and that is strongly baryon-dominated within $\sim$ 1 R$_{eff}$; and a more compact barely resolved component, $\sim 5 \times$ less massive than the northwest one. Both components are surrounded by an extended carbon-rich gas component out to $\sim$ 10 kpc.
        %\target is composed of two components and a carbon-rich gas surrounding the central system. We called these components \NW and \SE. \NW  is especially more extended and has a rising behavior in the position-velocity diagram. A clump-finding algorithm favors a single clump rather than two merging clumps. The component agrees with a rotating gaseous disk with a low dark matter fraction. 
        \item \textbf{Extended \cii emission:} Contrary to the \cii emission, the rest-UV and near-infrared continuum is compact, slightly elongated to the north-south direction. The $r_{[CII]}^{cir} = 1.6 \pm 0.3$ kpc is about 4 $\times$ greater than the UV effective extension. It extends up to 10 kpc compared to the average UV surface brightness radial profiles. It contributes up to $\sim$ 40\% of the \cii light. 
        \item \textbf{\cii extended emission in a merging system:} The upper and lower limits on the L$\rm_{FIR}$ and L$_{\cii}$/L$\rm_{FIR}$ values of the regions around the C05-NW and C05-SE components of CRISTAL-05 are consistent with those found in the overlap/shocked regions of the local merging systems, showing a enhancement in the \cii emission. Although we can not exclude the contribution of other mechanisms, the current kinematical status of the system points that the turbulence and shocks of interacting galaxies are the main drivers of the boost of the \cii emission in CRISTAL-05.

   \end{enumerate}

   Our results strongly exemplify the necessity of kpc-scale and pc-scales observations to completely dissect the internal dynamical state of galaxies 
   and to understand how it contributes to the metal enrichment circumgalactic medium of infant galaxies. Future observations with JWST and ALMA can help in stating the contribution of outflows in CRISTAL-05 by analyzing the hot ionized outflows, and the role of different physical mechanisms in the boost of \cii emission.
   
\begin{acknowledgements}

A. P. and M. A. acknowledge support from FONDECYT
grant 1211951, “ANID+PCI+INSTITUTO MAX PLANCK DE ASTRONO-
MIA MPG 190030”, “ANID+PCI+REDES 190194”,  “ANID scholarship 21221137”. 
RJA was supported by FONDECYT grant number 123171 and by the ANID BASAL project FB210003. RB acknowledges support from an STFC Ernest Rutherford Fellowship [grant number ST/T003596/1]. AF acknowledges support from the ERC Advanced Grant INTERSTELLAR H2020/740120. IDL, MP, and SvdG acknowledge funding support from ERC starting grant 851622 DustOrigin.  I.D.L and S.v.d.G acknowledges the support from the Flemish Fund for Scientific Research (FWO-Vlaanderen, projects G023821N and G0A1523N). R.I. is supported by Grants-in-Aid for Japan Society for the Promotion of Science (JSPS) Fellows (KAKENHI Grant Number 23KJ1006). E.d.C. gratefully acknowledges support from the Australian Research Council: Future Fellowship FT150100079, Discovery Project DP240100589, and the ARC Centre of Excellence for All Sky Astrophysics in 3 Dimensions (ASTRO 3D; project CE170100013).  TDS was supported by the Hellenic Foundation for Research and Innovation (HFRI) under the "2nd Call for HFRI Research Projects to support Faculty Members \& Researchers" (Project Number: 3382). K.T acknowledges support from JSPS KAKENHI Grant Number 23K03466. M. S. was financially supported by Becas-ANID scolarship \#21221511, and also acknowledges ANID BASAL project FB210003. TDS: The research project was supported by the Hellenic Foundation for Research and Innovation (HFRI) under the "2nd Call for HFRI Research Projects to support Faculty Members \& Researchers" (Project Number: 3382). HÜ gratefully acknowledges support by the Isaac Newton Trust and by the Kavli Foundation through a Newton-Kavli Junior Fellowship. RLD is supported by the Australian Research Council through the Discovery Early Career Researcher Award (DECRA) Fellowship DE240100136 funded by the Australian Government. M.K. was supported by the ANID BASAL project FB210003. MR acknowledges support from project PID2020-114414GB-100, financed by MCIN/AEI/10.13039/501100011033. 
This paper makes use of the following ALMA data: 
ADS/JAO.ALMA\#2017.1.00428.L, ADS/JAO.ALMA\#2021.1.00280.L,
ADS/JAO.ALMA\#2018.1.01359.S. ALMA is a partnership of ESO (representing its member states), NSF (USA) and NINS (Japan), together with
NRC (Canada), MOST and ASIAA (Taiwan), and KASI (Republic of Korea),
in cooperation with the Republic of Chile. The Joint ALMA Observatory
is operated by ESO, AUI/NRAO and NAOJ. 

\end{acknowledgements}

\bibliographystyle{aa}
\bibliography{aanda}

\newpage

\appendix

%\section{\cii moment-0 map}
%\label{appendix-moment0}
%In Figure~\ref{fig:appendix1}, we display the non-JvM-corrected \cii moment-0 maps for the combined datasets obtained using the \textit{tclean} task of CASA software using Briggs weighting, with robust = 0.5. The map represents the collapsed line cube, averaging the emission within the full-width tenth maximum centered at the \cii line frequency. 
%
%
%\begin{figure*}[h]
%    \centering
%     \includegraphics[width=0.55\textwidth]{Figures/A01-cii-moment-0-briggs.pdf}
%    \caption{ ALMA \cii moment-0 map (non-JvM corrected),  . The overlaid black contours correspond to the 3, 5, 7-$\sigma_{\cii}$ levels, where $\sigma_{\cii} = $ 0.02 Jy/beam km s$^{-1}$ is the rms noise level. The white ellipse indicates the beam size of $0.21\arcsec\times 0.18\arcsec$. }
%    \label{fig:appendix1}
%\end{figure*}

\section{Dynamical modelling}\label{appendix-dysmal}

\begin{figure*}[h]
    \centering
     \includegraphics[width=\textwidth]{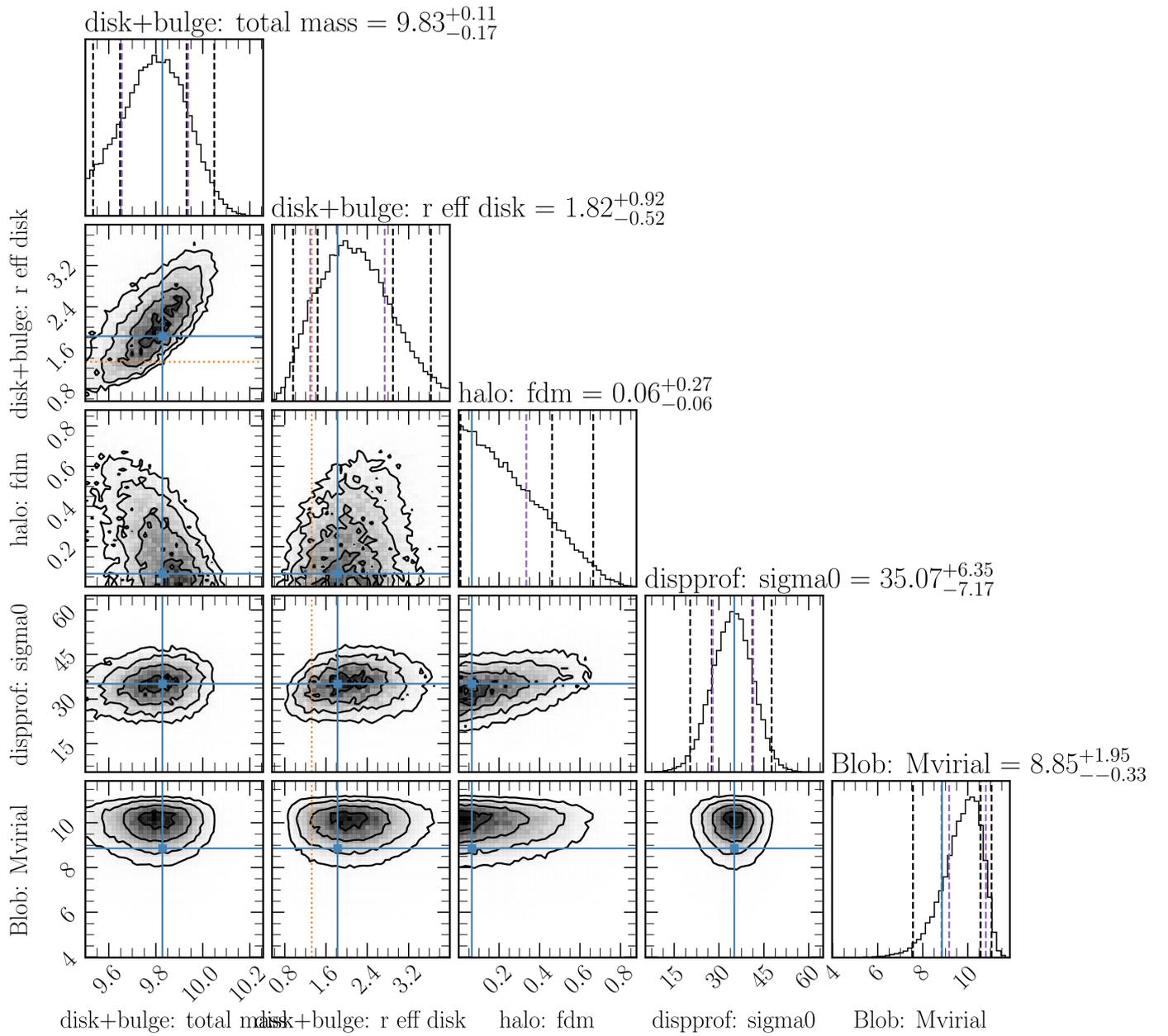}
    \caption{Posterior distributions for the kinematical modeling in C05-NW.}
    \label{fig:appendix2}
\end{figure*}

In Figures~\ref{fig:appendix2} we show the posterior distributions of the free parameters of the DYSMALpy routine. 

\section{CRISTAL-05 environment}\label{appendix-overdensity}

Our observations cover a primary beam of r$\rm_{PB}$ = 15\arcsec, which translates into a projected area of 25 Mpc$^{2}$. We report the detection of 3 galaxies in the continuum maps, with no \cii emission detected. Since the merger frequency is enhanced in overdensities, we explore the possibility that CRISTAL-05 is located close to the density peak of a protocluster environment. These galaxies are identified in the COSMOS2015 survey \citep{2016laigle} as 684856, 15086, and 686297, at distances of 51, 70, and 86 kpc, respectively. None of the sources have an available spectroscopic redshift, but COSMOS 2020 assigned a photometric redshift of z$_{phot}^{15086}$ = 1.91 and z$_{phot}^{686297}$ = 2.1. Target 684856 was not covered from COSMOS2020 but COSMOS 2015 found z$_{phot}^{684856}$ = 5.13. At this redshift, we did not find a \cii detection in the location of 684856. Therefore, we see no evidence that CRISTAL-05 is within a galaxy overdensity at z = 5.5 based on the existing data.

\end{document}

%% file: authors.tex
\author{A. Posses \inst{1} % CHECKED AFFILIATION 
  \and
  M. Aravena\inst{1}  % affiliation is right
  \and
  J. Gonz\'alez-L\'opez\inst{2,1,3} %nothing
  \and
  N. M. F\"orster Schreiber\inst{4}
  \and
  D. Liu\inst{4}
  \and
  L. Lee\inst{4}
  \and
  M. Solimano\inst{1} % affiliation and ack
  \and
  T. D\'iaz-Santos\inst{ 5, 6} % affiliation and ack
  \and
  R. J. Assef\inst{1} % checked affiliation and ack
  \and
  L. Barcos-Mu\~noz\inst{7,8} % checked affiliation and ack
  \and
  S. Bovino\inst{9, 10, 11} % MISSING
  \and
  R. A. A. Bowler\inst{12}  % AFFILIATION AND ACK
  \and
  G. Calistro Rivera\inst{13} % AFFILIATION AND CHECK
  \and
  E. da Cunha\inst{14,15} % affiliation and check
  \and
  R. L. Davies\inst{15, 18} 
  \and
  M. Killi\inst{1}
  \and
   I. De Looze\inst{16}
  \and
  A. Ferrara\inst{17}
  \and
  D. B. Fisher\inst{18, 12} % pensativa sobre  12
  \and
  R. Herrera-Camus\inst{9} 
  \and
  R. Ikeda\inst{19,20}%{21, 22}
  \and
  T. Lambert\inst{1}
  \and
  J. Li\inst{14}
  \and
  D. Lutz\inst{7}
  \and
  I. Mitsuhashi\inst{21,20}%{23, 22}
  \and
  M. Palla\inst{16, 10}
  \and
  M. Rela\~no\inst{22,23}%{24, 25} 
  \and
  J. Spilker\inst{24}
  \and
  T. Naab\inst{25}
  \and
  K. Tadaki\inst{26}
  \and
  K. Telikova\inst{1, 27}
  \and
  H. {{\"U}bler}\inst{28,29}
  \and
  S. van der Giessen\inst{16, 22,23}
  % I noticed that the affiliations for Stefan van der Giessen are not correct. It should be the one from me and the ones from Monica combined.
  \and
  V. Villanueva\inst{9}
}

\institute{Instituto de Estudios Astrof\'isicos, Facultad de Ingenier\'ia y Ciencias, %
  Universidad Diego Portales, Av.  Ej\'ercito Libertador 441, Santiago, Chile [C\'odigo Postal 8370191]
  \email{ana.posses@mail.udp.cl} 
  \and
  %2
   Instituto de Astrof\'isica, Facultad de F\'isica, %
  Pontiﬁcia Universidad Cat\'olica de Chile, Santiago 7820436, Chile
  \and
  %3
  Las Campanas Observatory, Carnegie Institution of Washington, %
  Casilla 601, La Serena, Chile
  \and
  %4
  Max-Planck-Institut f\"ur Extraterrestiche Physik (MPE), %
  Giessen-bachstr., 85748, Garching, Germany
  \and
  %5
  Institute of Astrophysics, Foundation for Research and Technology-Hellas (FORTH), Heraklion, GR-70013, Greece
  \and
  %6
  School of Sciences, European University Cyprus, Diogenes street, Engomi, 1516 Nicosia, Cyprus
  \and
  %7
  National Radio Astronomy Observatory, 520 Edgemont Road, Charlottesville, VA 22903, USA
  \and
  %8
  Department of Astronomy, University of Virginia, 530 McCormick Road, Charlottesville, VA 22903, USA
  \and
  %9
  Departamento de Astronomía, Facultad Ciencias Físicas y Matemáticas, Universidad de Concepción, Av. Esteban Iturra s/n Barrio Universitario, Casilla 160, Concepción, Chile
  \and
  %10
  INAF, Istituto di Radioastronomia — Italian node of the ALMA Regional Centre (It-ARC), Via Gobetti 101, 40129 Bologna, Italy
  \and
  %11
  Dipartimento di Chimica, Università degli Studi di Roma “La Sapienza”, P.le Aldo Moro 5, 00185 Roma, Italy
  \and
  %12
  Jodrell Bank Centre for Astrophysics, Department of Physics and Astronomy, School of Natural Sciences, The University of Manchester, Manchester, M13 9PL, UK
  \and
  %13
   European Southern Observatory, Karl-Schwarzschild-Str. 2, D-85748, %
   Garching, Germany
   \and
   %14
   International Centre for Radio Astronomy Research, University of Western Australia, 35 Stirling Hwy, Crawley 26WA 6009, Australia
   \and
   %15
  ARC Centre of Excellence for All Sky Astrophysics in 3 Dimensions (ASTRO 3D), Australia
  \and
  %16
  Sterrenkundig Observatorium, Ghent University, Krijgslaan%
  281-S9, B-9000 Ghent, Belgium
  \and
  %17
  Scuola Normale Superiore, Piazza dei Cavalieri 7, 50126 Pisa, Italy
  \and
  %18
  Centre for Astrophysics and Supercomputing, Swinburne Univ. %
  of Technology, PO Box 218, Hawthorn, VIC, 3122, Australia
  \and
  %19
  Department of Astronomical Science, SOKENDAI (The Graduate University for Advanced Studies), Mitaka, Tokyo 181-8588, Japan
  \and
  %20
   National Astronomical Observatory of Japan, 2-21-1 Osawa, %
   Mitaka, Tokyo 181-8588, Japan
   \and
   %21
   Department of Astronomy, The University of Tokyo, 7-3-1 %
  Hongo, Bunkyo, Tokyo 113-0033, Japan
  \and
  %22
  Dept. Fisica Teorica y del Cosmos, Universidad de Granada, %
  Granada, Spain
  \and
  %23
  Instituto Universitario Carlos I de F\'{i}sica Te\'{o}rica %
  y Computacional, Universidad de Granada, %
  E-18071 Granada, Spain
  \and
  %24
  Department of Physics and Astronomy and George P. and %
  Cynthia Woods Mitchell Institute for Fundamental Physics %
  and Astronomy, Texas A\&M University, College Station, TX, USA
  \and
  %25
  Max-Planck Institute for Astrophysics, %
  Karl Schwarzschildstrasse 1, 85748, Garching, Germany
  \and
  %26
  Faculty of Engineering, Hokkai-Gakuen University, Toyohira-ku, %
  Sapporo 062-8605, Japan
  \and
  %27
  Ioffe Institute, 26 Politekhnicheskaya, St. Petersburg, 194021, Russia
  \and
  %28
  Cavendish Laboratory, University of Cambridge, 19 J.J. %
  Thomson Avenue, Cambridge, CB3 0HE, UK
  \and
  %29
  Kavli Institute for Cosmology, University of Cambridge, %
  Madingley Road, Cambridge, CB3 0HA, UK
  }